\documentclass[]{aastex63}

\received{January 05, 2023}
\revised{February 16, 2023}
\accepted{February 21, 2023}
\submitjournal{AJ}
\shorttitle{Galactic model parameters and space density of CV}
\shortauthors{Canbay et al.}
\usepackage{xcolor}
\usepackage{rotating}
\begin{document}

\title{Galactic Model Parameters and Space Density of Cataclysmic Variables in 
Gaia Era: New Constraints to Population Models}

\correspondingauthor{Remziye Canbay}
\email{rmzycnby@gmail.com}

\author[0000-0003-2575-9892]{Remziye Canbay}
\affiliation{Istanbul University, Institute of Graduate Studies in Science, 
Programme of Astronomy and Space Sciences, 34116, Beyaz{\i}t, Istanbul, Turkey}

\author[0000-0003-3510-1509]{Sel\c{c}uk Bilir}
\affiliation{Istanbul University, Faculty of Science, Department of Astronomy 
and Space Sciences, 34119, Beyaz\i t, Istanbul, Turkey}

\author[0000-0003-1399-5804]{Aykut \"Ozd\"onmez}
\affiliation{Atat\"urk University, Faculty of Science, Department of Astronomy 
and Space Sciences, 25240, Erzurum, Turkey}

\author[0000-0002-0688-1983]{Tansel Ak}
\affiliation{Istanbul University, Faculty of Science, Department of Astronomy 
and Space Sciences, 34119, Beyaz\i t, Istanbul, Turkey} 

\begin{abstract}
\noindent
The spatial distribution, Galactic model parameters and luminosity function of 
cataclysmic variables (CVs) are established using re-estimated trigonometric 
parallaxes of {\it Gaia} DR3. The data sample of 1,587 CVs in 
this study is claimed to be suitable for Galactic model parameter estimation  
as the distances are based on trigonometric parallaxes and the {\it Gaia} 
DR3 photometric completeness limits were taken into account when the sample 
was created. According to the analysis, the scale height of All CVs increases 
from 248$\pm$2 to 430$\pm$4 pc towards shorter periods near the lower limit of 
the period gap and suddenly drops to 300$\pm$2 pc for the shortest orbital 
period CVs. The exponential scale heights of All CVs and magnetic systems 
are found to be 375$\pm$2 and 281$\pm$3 pc, respectively, considerably larger 
than those suggested in previous observational studies. The local space density 
of All CVs and magnetic systems in the sample are $6.8^{+1.3}_{-1.1}\times$10$^{-6}$ 
and $2.1^{+0.5}_{-0.4}\times10^{-6}$ pc$^{-3}$, respectively.  
 Our measurements strengthen the 1-2 order of magnitude discrepancy 
between CV space densities predicted by population synthesis models and 
observations. It is likely that this discrepancy is due to objects 
undetected by CV surveys, such as the systems with very low $\dot{M}$ and   
the ones in the period gap. The comparisons of the luminosity function of white 
dwarfs with the luminosity function of All CVs in this study show that 500 times 
the luminosity function of CVs fits very well to the luminosity function of 
white dwarfs. We conclude that the estimations and data sample in this study 
can be confidently used in further analysis of CVs.
\end{abstract}

\section{Introduction}

\label{sec:introduction}
Cataclysmic variables (CVs) are short-period semi-detached binary stars. A cataclysmic 
variable’s primary component is a white dwarf which is accreting matter from a Roche-lobe 
filling low-mass main-sequence star, the secondary component, via 
a gas stream. Since the matter stream has a high angular momentum and the primary 
star is small, an accretion disc surrounding the white dwarf is created. A bright 
spot is also formed where the matter stream impacts the disc. Magnetised white 
dwarfs in CVs have accretion columns instead of discs to transfer matter from 
the secondary component \citep*{War95,Hel01,Knig11a,Knig11b}. 

The standard formation and evolution scenario developed for CVs is concentrated 
on the explanation of the features seen in the orbital period distribution of 
these systems, since the most precisely determined parameter of a CV is its 
orbital period. The sharp cut-off at about 80 min \citep{Wil05,Gan09}, period 
minimum, and the period gap between roughly 2 and 3 h \citep{King88,Knig11b} 
are the most striking features of the CV period distribution. The standard 
theory successfully explains these features as its main predictions are supported 
by observations. However, there are still some observational properties  
to be explained, for example (1) Predicted and observed fractions of CVs above 
and below the period gap \citep{deKool92,Kolb93,How01,Gan09,McAl19} are not in 
agreement. Standard CV population studies predict that more than 90\% of CVs must 
be located below the period gap, while observations imply almost equal numbers of 
CVs below and above the gap. Although sky surveys like the Sloan Digital Sky Survey 
\citep[SDSS;][]{Szk02,Szk03,Szk04,Szk05,Szk06,Szk07,Szk11} revealed a larger 
population of short-period CVs, the disagreement still remains. (2) The fraction 
of observed post-period minimum CVs, period bouncers, is much smaller than that 
predicted by the population models based on the standard theory 
\citep{Pat05,Unda08,LitF08,Pat11,Kato15,Kato16,McAl17,Neus17,Pala18}. \citet{McAl19} 
found that 30\% of donor stars in a sample of 225 CVs are likely to be brown dwarfs 
in period bouncers. However, only 5\% of volume-limited sample of CVs in 
\citet{Pala20} included period bouncers. (3) Although the standard theory predicts 
an orbital period minimum of about 65-70 min \citep[see][and references therein]{Kalo16}, 
observed values are about 76-82 min \citep{Knig11b,McAl19}. (4) The observed white 
dwarf masses in CVs have been significantly larger than those of single white 
dwarfs \citep[see][and references therein]{ZS20}. In addition, white dwarf mass in CVs 
does not change with orbital period \citep{McAl19}. (5) Population studies based on 
the standard evolutionary model predict space densities 1-2 orders of magnitude larger 
than observed values \citep{ZS20}. Although additional angular momentum loss 
mechanisms and models \citep*{Pat98,Knig11b,SZW16,Pala17,ZS17,ZS20,Bel18,LiLi19,Metz21,Sark22} 
were suggested to solve the disagreements, the only published, self-consistent simulations 
of CV evolution were performed by \citet{Hill20}, whose multi-Gyr models of novae 
take into account every nova eruption's thermonuclear runaway, mass and angular 
momentum loses, feedback due to irradiation and variable mass transfer rate ($\dot{M}$), 
and orbital size and period changes. \citet{Hill20} reproduced the observed range of 
mass transfer rates at a given orbital period, with large and cyclic Kyr-Myr 
timescale changes. It should be noted that the magnetic systems may have different 
evolutionary scenarios from non-magnetic CVs \citep[see references in][]{Bel20}.

Depending on the completeness of the samples, reliable observational constraints 
can be obtained from the stellar statistics \citep*{Ak08,Ozdn15}, and a proposed 
evolutionary scheme must also be in agreement with the data obtained from stellar 
statistics \citep{Duer84}. In this respect, Galactic model parameters and the space 
density of a group of objects are key parameters to constrain and to test population 
models based on evolutionary schemes. A wide range of observational results is 
remarkable, while the predicted space densities are systematically 1-2 order of 
magnitude larger than those derived from observations. For example, previous CV 
population synthesis models predicted space densities $10^{-5}$-$10^{-4}$ pc$^{-3}$  
\citep{RB86,deKool92,Kolb93,Pol96,Wil05,Wil07,GN15,Bel18}, while observations 
indicated $10^{-7}$-$10^{-4}$ pc$^{-3}$  
\citep*{War74,Pat84,Pat98,TB98,Rin93,Schw02,Arau05,Pret07,PKK07,Ak08,Revn08,PK12,PKS13,Schw18}. 
In a recent study, \citet{Pala20} measured very precise space densities 
of $4.8^{+0.6}_{-0.9}\times 10^{-6}$ and $1.2^{+0.4}_{-0.5} \times10^{-6}$ pc$^{-3}$ 
for All CVs and magnetic CVs (mCVs), respectively, using the European Space Agency’s 
(ESA) {\it Gaia} Data Release 2 \citep[{\it Gaia} DR2;][]{Gaia18}. They assumed a scale 
height of 280 pc for their analysis and the sample was composed of only 42 objects 
within 150 pc from the Sun. They assumed that this restriction reduces the uncertainties 
in the derived space densities related to the unknown age and scale height of the 
CV population and the uncertainties from astrometric solutions. \citet{Bel20} 
claimed that these are the most reliable observational space density estimates ever 
found. They also concluded that the space densities given in \citet{Pala20} are in 
very good agreement with their predicted values if potential period bouncers are 
excluded from the space density estimation. However, it should be noted that 
\citet{BelSch20} performed binary population models using an up-to-date version 
of the BSE code \citep{HTP02} and found that their model fails to explain some 
observational properties of magnetic CVs. Thus, the agreement between the space 
density measurements of \citet{Pala20} and predictions by \citet{Bel20} does not 
mean that the previous population synthesis models based on the standard evolution 
and formation theory are wrong. It is likely that the space densities of 
$10^{-5}$-$10^{-4}$ pc$^{-3}$ proposed by previous CV population synthesis models 
are correct and the current observational measurements suffer from incomplete 
sky surveys as the surveys are probably missing the most of very low $\dot{M}$ 
systems, CVs in the orbital period gap, whose lifetime is predicted to be longer 
\citep{Hill20}, and the most of period bouncers.

Besides these studies, observational Galactic model parameters, luminosity functions 
and space densities of CVs, i.e. intrinsic properties of Galactic CV population, 
should still be determined from suitable data by using the methods of observational 
Galactic structure studies \citep*{Karaali04,Biletal06a,Biletal06b,Biletal06c,Biletal08,
Karaali07,Cabrera07}, as the number of systems with reliable distance estimates is high 
enough to use these methods. Sky surveys such as {\it Gaia} 
\citep{Gaia16,Gaia18,Gaia21a,Gaia22} and SDSS \citep{Szk02,Szk03,Szk04,Szk05,Szk06,Szk07,Szk11} 
determined faint systems and presented reliable distances of CVs. Using these data, it 
is possible to decrease the selection effects that may be strong for faint systems. 
These data and methods can allow us to determine observational constraints for 
evolutionary models of CVs and mCVs. 

In this study, we use ESA's {\it Gaia} Data Release 3 
\citep[{\it Gaia} DR3;][]{Gaia22} and \citet{Bai21} to obtain reliable 
distances for a CV sample. We analyse the spatial distribution of CVs and discuss the 
completeness of the CV sample. Then, we use this sample to estimate the Galactic 
model parameters and space densities of All CVs and mCVs with scale-heights 
obtained from exponential and ${\rm sech}^2$ functions fitted to $z$-histograms.       

\section{Data}
In order to construct our CV sample, we collected CVs from AAVSO's International 
Variable Star Index\footnote{www.aavso.org/vsx/} database. Their equatorial and Galactic 
coordinates were taken from SIMBAD\footnote{https://simbad.u-strasbg.fr/simbad/sim-fid} 
database. We also included CV’s found in the previous studies 
\citep[e.g.][]{Hoff18, Hal18, Szk18, Ber19, Kato19, Yu19, Bel20, Kato20, Schw20} in this 
sample. Orbital periods were mainly taken from \citet{RK03} and \citet{Dow01}. Superhump 
periods of objects, whose orbital periods are unknown, were assumed to be their orbital 
periods \citep[][and references therein]{Kato20}. The number of systems is 10,852 in 
this very rough sample. We ignored the objects classified as non-mCVs and removed objects 
for which 
trigonometric parallax measurements are not present in {\it Gaia} DR3. All the objects in 
the preliminary sample were checked with respect to equatorial coordinates in order to 
avoid from duplication. In order to prevent the misidentification of CVs due to adjacent 
objects, we also checked each object’s {\it Gaia} position individually using 
Aladin\footnote{https://aladin.u-strasbg.fr/AladinLite/} and removed misidentified objects 
from the database. To ensure a robust match, we crossmatched each object in the catalogue 
to a $15^{''}$ radius in {\it Gaia} DR3, propagated the subsets of {\it Gaia} DR3 $15^{''}$ 
crossmatches to the J2000 epoch using proper motion, and made a final crossmatch at J2000 
with a radius of $2^{''}$, given the precision of the catalogue. The 132 CVs in the 
directions of the globular clusters were excluded from the statistics to avoid mismatching 
with the data in the {\it Gaia} catalogue. As a result of this selection process, 
{\it Gaia} photometric and astrometric data of 5,621 CVs were obtained. We showed 
the matching procedure for nine objects as an example. The results for matching these objects 
at different $G$ apparent magnitudes from the {\it Gaia} DR3 catalogue with panSTARR $g$ 
images on Aladin\footnote{http://aladin.cds.unistra.fr/aladin.gml} are shown in Figure 1.

%FIGURE 01
\begin{figure*}
\begin{center}
\includegraphics[scale=0.80]{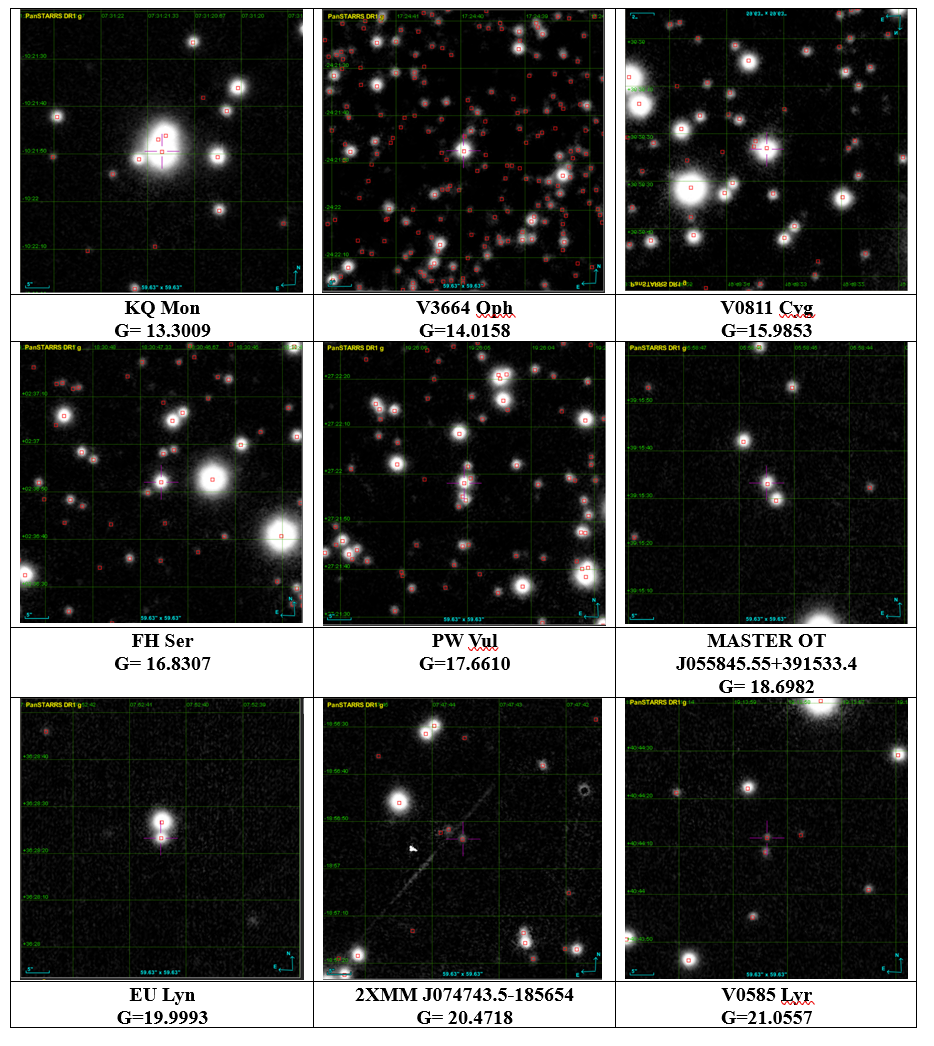}
\caption{Location of nine CVs with different $G$ apparent magnitudes selected from 
the {\it Gaia} DR3 catalogue are shown on panSTARR $g$ images using Aladin. Target objects are 
located in the centres of the images.}
\end{center}
\end{figure*}

As we need a precise data sample to extract Galactic model parameters of CVs, we 
also made strict cuts on quality flags, even though these cuts remove numerous systems 
from the sample. We retain matches with \texttt{phot\_g\_mean\_flux\_over\_error}, 
$f_{\rm G}/{\delta f_{\rm G}}, > 8$, more than eight “good” astrometric observations, 
as characterized by {\it Gaia} DR3, \texttt{astrometric\_excess\_noise<2}, and 
$\varpi>0.1$ mas. The remaining sample includes 4,149 CVs. In the sample, magnetic CVs 
classified as DQ Her (Intermediate polars) or AM Her type (Polars) objects are 
indicated as mCV, remaining systems as CV. Number of magnetic CVs is only 205 in 
this sample. Although the sample includes 4,149 objects classified 
as CV, we know orbital periods only for 1,187 of them.

We used {\it Gaia} DR3 data \citep{Gaia22} to obtain the distances of CVs in our catalogue. 
In order to do this, we matched our catalogue with {\it Gaia} DR3 catalogue and 
found {\it Gaia} ID for each CV. It is possible to estimate the distances of CVs by simply 
inverting their trigonometric parallaxes taken from {\it Gaia} DR3 catalogue. 
However, \citet{Bai18,Bai21} indicated that the nonlinearity of the transformation 
and the asymmetry of the resulting probability distribution must be taken into account, 
and they re-estimated the {\it Gaia} EDR3 \citep{Gaia21a} parallaxes. In such 
a re-estimation, we do not expect too different distance estimates from {\it Gaia} DR3 data 
and \citet{Bai21}'s approximation for a certain system. Thus, we matched our catalogue 
with the catalogue of \citet{Bai21} using {\it Gaia} DR3 IDs to obtain precise distances 
and distance errors of CVs and compared distances of CVs estimated from {\it Gaia} DR3 
and \citet{Bai21} in Figure 2, where different $G$ apparent magnitude intervals are 
represented by coloured symbols. This comparison shows that there is a considerable 
scatter especially for fainter systems, while the scatter is much less for CVs with 
$G\leq 18.5$ mag. Thus, we limited our CV sample to the systems for which 
$G\leq 18.5$ mag, which is set as the faint limit of the sample. We set the bright 
limit as $G=9$ mag, since there is no brighter object in our sample. The final sample 
comprises of CVs with $9\leq G \leq 18.5$ mag and includes 1,714 CVs, 767 of them with 
known orbital periods. The relative distance error of the sample is less than 1.66 and 
the median value is 0.06.

%FIGURE 02
\begin{figure*}
\begin{center}
\includegraphics[width=0.8 \textwidth]{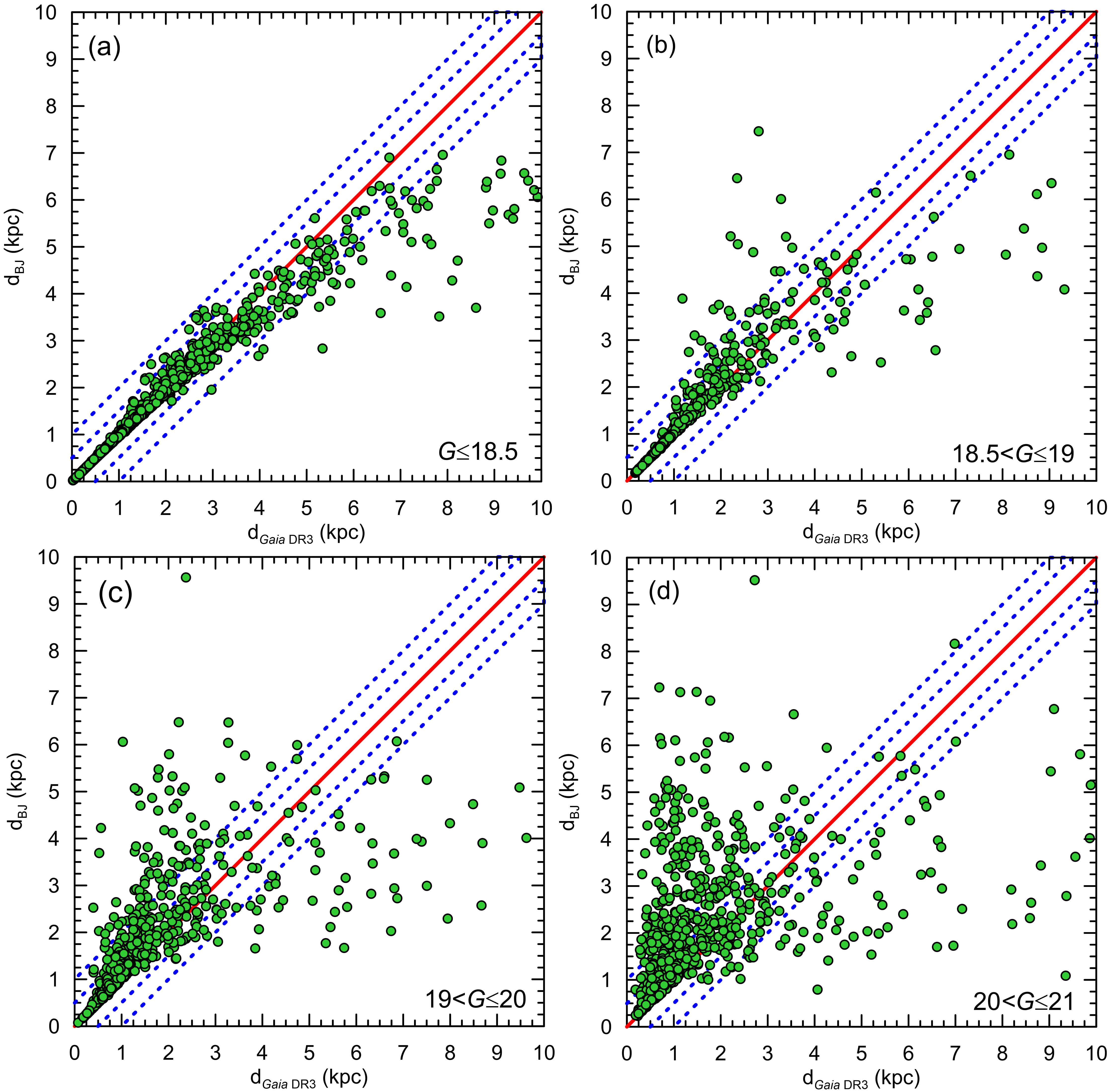}
\caption{Comparison of CV distances obtained from {\it Gaia} DR3 and \citet{Bai21} 
($d_{\rm BJ}$). Different $G$ apparent magnitude intervals are shown in 
panels (a), (b), (c) and (d). The red line represents the one-to-one line and blue 
dashed lines 500 and 1000 pc distances from the red line.}
\end{center}
\end{figure*}

Although the number of CVs in the sample is the largest ever used in similar 
analyses, it is still questionable if this sample is sufficiently large to be 
representative of the entire CV population and subject to magnitude-related selection 
effects. Since the standard theory predicts that most CVs should be intrinsically 
faint objects, it seems that apparent magnitude limits of surveys are one of the 
strongest selection effects. Therefore, the completeness limits of the data must be
taken into account in a study based on stellar statistics. 

In order to set completeness limits to the sample, we first obtained interstellar 
absorption in $V$-band $A_{\rm V}$ for CVs in the sample by using 
MWDUST\footnote{https://github.com/jobovy/mwdust} code which produces two and 
three-dimensional Galactic dust map \citep{Bovy16}, based on \citet{Schletal98}'s 
dust maps as re-calibrated by \citet{SchFinketal11}. 
MWDUST code can provide two and three-dimensional data according to 
Galactic coordinates from the Sun to all directions until the edges of our Galaxy 
contributed by the Galactic dust. Since the distances of systems are well-known 
from trigonometric parallaxes, we preferred to use two-dimensional data and 
showed the rough absorption value in $V$-band according to Galactic latitude ($b$) 
and longitude ($l$) toward the direction of an object by $A_{\infty}(V)$, 
practically means up to infinite, but actually up to the edge of the Galaxy. 

The total absorption in $V$-band for the distance $d$ to the star is calculated as 
following \citep{BS80}

\begin{equation}
A_{\rm d}(V)=A_{\infty}(V)\Biggl[1-\exp\Biggl(\frac{-\mid d\times\sin b\mid}{H}\Biggr)\Biggr],
\end{equation}
here $H$ is the scaleheight for the interstellar dust which is adopted as 
125 pc \citep{Mars06}. As the distance of the system ($d_{\rm BJ}$, hereafter 
also denoted as $d$) $d$ is known from \citet{Bai21}, the total absorption for 
the system $A_{\rm d}(V)$ could be estimated. We obtained the colour excess 
$E_{\rm d}(B-V)$ for each CV in the sample using $E_{\rm d}(B-V)=A_{\rm d}(V)/3.1$. 
The total absorptions in $G$, $G_{\rm BP}$, and $G_{\rm RP}$ bands were obtained by using 
relations given as follows: 
\begin{eqnarray}
A(G)=0.83627\times 3.1 E_{\rm d}(B-V), \nonumber \\ 
A(G_{\rm BP})=1.08337\times 3.1 E_{\rm d}(B-V), \\ 
A(G_{\rm RP})=0.63439\times 3.1 E_{\rm d}(B-V). \nonumber\\ 
\nonumber
\end{eqnarray}
The selective absorption coefficients in Equation (2) were taken from \citet{Cardelli89}. Once 
we know the total absorption values, we could calculate the de-reddened apparent 
magnitudes in $G$, $BP$ and $RP$-bands, $G_{0}$, $G_{\rm BP}$ and $G_{\rm RP}$, respectively. 
The absolute magnitudes $M_{\rm G}$ of CVs were calculated using the distance modulus 
formula $G_{\rm 0}-M_{\rm G} = 5\times \log(d_{\rm BJ})-5$, where $d_{\rm BJ}$ is the 
distance obtained from \citet{Bai21}. The absolute magnitudes $M_{\rm G}$ of CVs in the 
preliminary sample against their distances are shown in Figure 3.

%FIGURE 03
\begin{figure}
\begin{center}
\includegraphics[width=0.6 \textwidth]{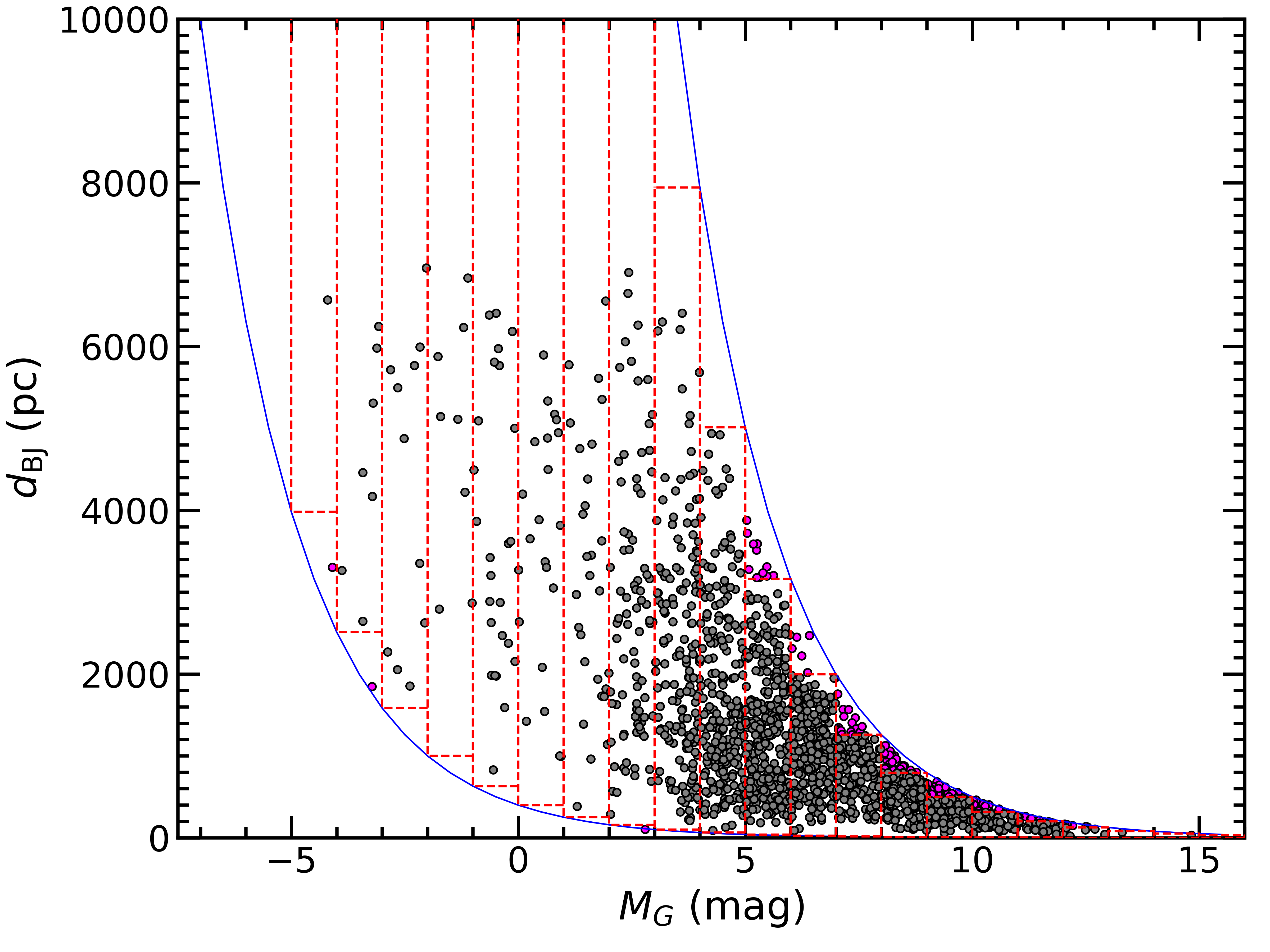}
\caption{The absolute magnitudes $M_{\rm G}$ of CVs in the preliminary sample 
against their distances $d_{\rm BJ}$ obtained from \citet{Bai21}. Red dashed lines show 
distances estimated from the bright ($G=9$ mag) and  faint ($G=18.5$ mag) limiting 
apparent magnitudes for absolute magnitude intervals of 1 mag. Blue solid 
lines correspond to the bright and faint apparent magnitude limits.}\label{Fig2}
\end{center}
\end{figure}

Red dashed lines in Figure 3 show distances estimated from the bright and faint brightness
limits ($9 \leq G \leq 18.5$ mag) for absolute magnitude intervals of 1 mag. These boxes 
limited by red dashed lines define the completeness limits of the data for certain absolute 
magnitude intervals. We removed systems beyond (out of the boxes defined by red dashed lines 
in Figure 3) these limiting magnitudes from the CV sample in order to obtain a complete 
catalogue in a certain volume with the Sun in its centre. We found that the majority of systems 
beyond 4 kpc in Figure 3 are discovered by SDSS. The final sample includes 1,587 CVs, 704 
of them with known orbital periods. There are only 124 mCVs in this sample, 117 of them with 
a known orbital period. Analyses in this paper were performed using this final sample. 

The final sample is given in Table 1 including equatorial coordinates 
$(\alpha, \delta)_{\rm J2000}$, object groups (magnetic (mCVs) or non-magnetic (non-mCVs)), 
orbital periods ($P_{\rm orb}$), {\it Gaia} DR3 trigonometric parallaxes 
($\overline{\omega}$) and relative parallax errors 
($\sigma_{\overline{\omega}}/{\overline{\omega}}$), proper motions 
($\mu_{\alpha}\cos\delta$, $\mu_{\delta}$), $G$-band apparent magnitudes 
from {\it Gaia} DR3. Distances $d_{\rm BJ}$ (also denoted as $d$) in Table 1 
were taken from \citet{Bai21}. This sample is the largest ever used in similar analyses 
of CVs. Besides, it includes the most reliable distance information for these systems. 
Objects, that are not classified as magnetic system in the literature, are 
denoted as CV in our catalogue.

%TABLE 1
\begin{table*}
\setlength{\tabcolsep}{2.5pt}
\begin{center}
\tiny{
\caption{The data sample. The object name, equatorial coordinates 
$(\alpha, \delta)_{\rm J2000}$, orbital periods $P_{\rm orb}$ and object groups 
(magnetic (mCV) or non-magnetic (non-mCVs)). $G$ denotes $G$-band apparent magnitudes, 
$\overline{\omega}$ trigonometric parallaxes, 
$\sigma_{\overline{\omega}}/{\overline{\omega}}$ relative parallax errors,   
$\mu_{\alpha}\cos\delta$ and $\mu_{\delta}$ proper motion components from 
{\it Gaia} DR3. Distance $d_{\rm BJ}$ is calculated using the system parallax 
in \citet{Bai21}. First and the last two rows of Table 1 is given here. 
The table can be obtained electronically.}
\begin{tabular}{llccccccccccc}
\hline
ID &  Star Name   &  $\alpha$ & $\delta$ &  Object &  $P_{\rm orb}$ & $\overline{\omega}$  & $\sigma_{\overline{\omega}}/{\overline{\omega}}$ & $\mu_{\alpha}\cos\delta$ & $\mu_{\delta}$ & $G$ & $d_{\rm BJ}$ & Ref\\ 
     &        & (hh:mm:ss) & (dd:mm:ss) &    groups      &  (d) & (mas) &  &  (mas yr$^{-1}$)  & (mas yr$^{-1}$) & (mag) & (pc) & \\
 \hline
    0001 & 2XMM J000134.1+625008 & 00 01 34.18 & +62 50 07.40 &  non-mCVs & .... & 0.2519 & 0.0968 &  2.268$\pm$0.091 & 0.486$\pm$0.0941 & 17.9224$\pm$0.0043 & 4133$\pm$1764 & ....\\
    0002 & EF Tuc & 00 01 55.10 & -67 07 43.3 & non-mCVs & 0.145 & 0.7120 & 0.0203 &  13.348$\pm$0.020 & -1.455$\pm$0.0230 & 14.8480$\pm$0.0196 & 1330$\pm$36 & (01)\\
  .... & .... & .... & .... & .... & ....& .... & .... & .... & .... & .... & .... & ....\\
    .... & .... & .... & .... & .... & .... & .... & .... & .... & .... & .... & .... & .... \\
      .... & .... & .... & .... & .... & .... & .... & .... & .... & .... & .... & .... & .... \\
    1713 & 2dFGRS TGS429Z114 & 23 50 32.90 & -31 25 48.5 & non-mCVs & - & 0.2672 & 0.0585 &  -2.237$\pm$0.043 & -3.851$\pm$0.045 & 16.6399$\pm$0.0028 & 3338$\pm$715 & ....\\
  1714 & BC Cas & 23 51 17.45 & +60 18 10.0 & non-mCVs & - & 0.5360 & 0.0588 &  -0.861$\pm$0.059 & -2.52$\pm$0.06 & 17.2256$\pm$0.0095 & 1831$\pm$198 & ....\\
\hline

\end{tabular}  
}
\end{center}
(01) \citet{Dow01}, (02) \citet{Hardy17}, (03) \citet{RK03}, (04) \citet{Szk13}, (05) \citet{Gonz13}, (06) \citet{Steiner07}, (07) \citet{Paterson19}, (08) \citet{Kozhevnikov19}, (09) \citet{Thorstensen17}, (10) \citet{Watson06}, (11) \citet{Wils11}, (12) \citet{Thorstensen16}, (13) \citet{Rude12}, (14) \citet{Nesci19}, (15) \citet{Pat03}, (16) \citet{Hamilton18}, (17) \citet{Sterken07}, (18) \citet{Thorstensen20}, (19) \citet{Coppejans16}, (20) \citet{Plavchan08}, (21) \citet{Copperwheat11}, (22) \citet{Shafter06}, (23) \citet{Bruch19}, (24) \citet{Masci19}, (25) \citet{Hummerich14}, (26) \citet{McA15}, (27) \citet{Thorstensen20a}, (28) \citet{Kato15}, (29) \citet{Kozhevnikov14}, (30) \citet{Thorstensen03}, (31) \citet{Drake14}, (32) \citet{Thorstensen10}, (33) \citet{Svhaefer21}, (34) \citet{Ruiz-Carmona20}, (35) \citet{Kato09}, (36) \citet{Kozhevnikov18}, (37) \citet{Pala20}, (38) \citet{Hal15}, (39) \citet{Alfonso12}, (40) \citet{Rod12}, (41) \citet{Kato17}, (42) \citet{Kozhevnikov03}, (43) \citet{Hal18}, (44) \citet{Worpel18}, (45) \citet{Kato20a}, (46) \citet{Rod05}, (47) \citet{Thorstensen12}, (48) \citet{Joshi20}, (49) \citet{Dai20}, (50) \citet{Ber19}, (51) \citet{Bruch17}, (52) \citet{Bel20}, (53) \citet{McAl19}, (54) \citet{Mason13}, (55) \citet{Uemura10}, (56) \citet{Beuermann21}, (57) \citet{Samus17}, (58) \citet{Kato20}, (59) \citet{Pal20}, (60) \citet{Tappert13}, (61) \citet{Vogt18}, (62) \citet{Rutkowski11}, (63) \citet{Avil20}, (64) \citet{Chen20}, (65) \citet{Koen95}, (66) \citet{Gabdeev19}, (67) \citet{Rin12}, (68) \citet{Han16}, (69) \citet{Mukai09}, (70) \citet{Hambsch14}, (71) \citet{Kato13}, (72) \citet{Bond18}, (73) \citet{Zubareva11}, (74) \citet{Yu19}, (75) \citet{Sheets07}, (76) \citet{Breus19}, (77) \citet{Gasque19}, (78) \citet{Myers17}, (79) \citet{Kozhevnikov17}, (80) \citet{Rude12a}, (81) \citet{Altan19}, (82) \citet{Weil18}

\end{table*}

In Figure 4a, we show the distance histogram of CVs in the final sample which is 
limited using bright and faint limiting magnitudes in $G$-band, 9 and 18.5 mag, 
respectively. The cumulative distribution of CV distances is presented in Figure 4b. 
Relative distance errors $\sigma_{d_{\rm BJ}}/{d_{\rm BJ}}$ obtained 
from \citet{Bai21} are also shown in Figure 4c. As expected, the 
relative errors are increasing with the distance. Relative distance errors 
are good indicators of the precision of the distance measurements, which is 
very important in our study. While relative errors for 66\% of 
the systems in the final sample is 
$\sigma_{d_{\rm BJ}}/{d_{\rm BJ}}$ $\leq$ 0.10, 88\% of them 
have relative distance errors of $\sigma_{d_{\rm BJ}}/{d_{\rm BJ}}$ $\leq$ 0.25. 
These values show that reliable constraints to population models of CVs can be obtained 
from the sample in this study. 

%FIGURE 4
\begin{figure}
\begin{center}
\includegraphics[scale=0.15]{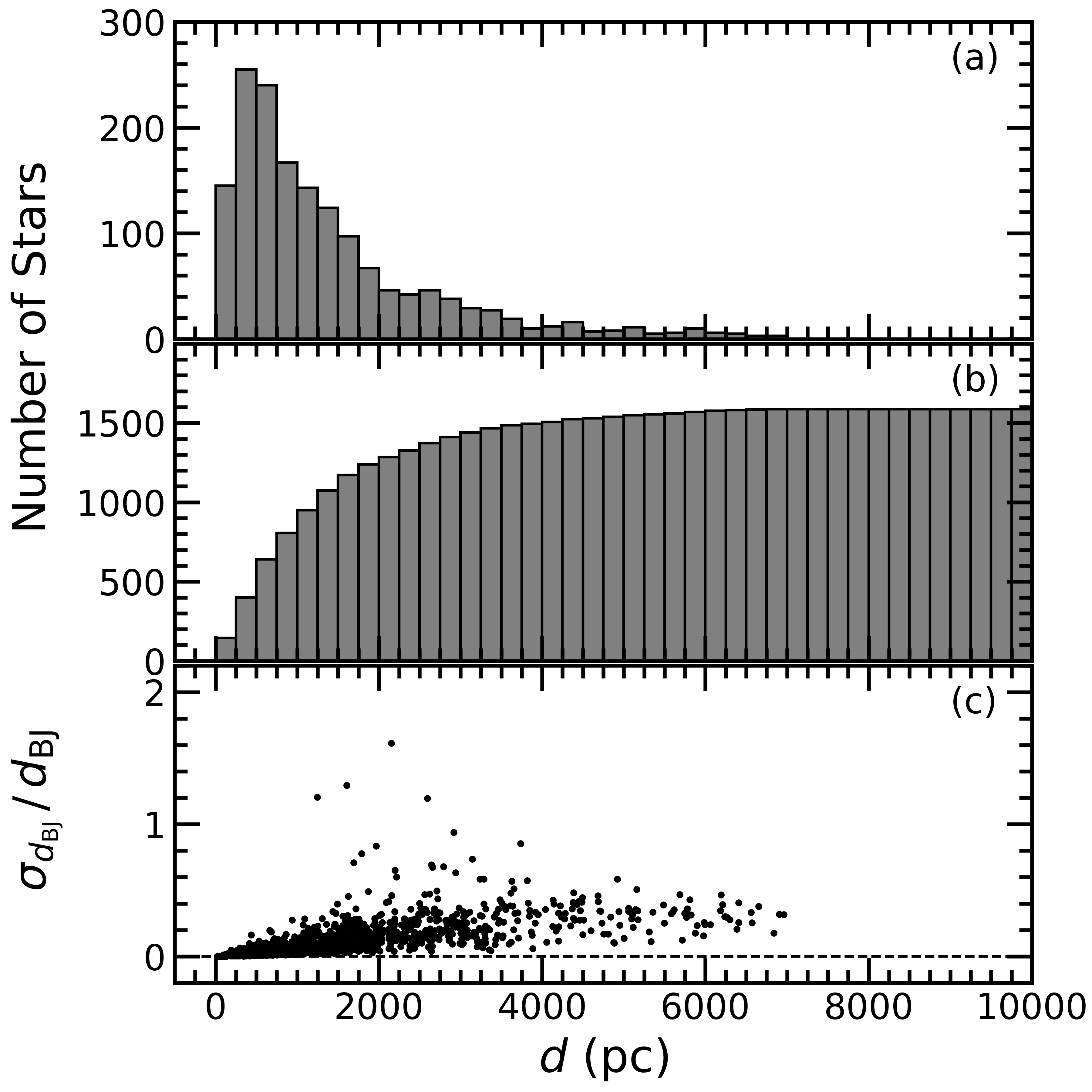}
\caption{Distance histogram of CVs in the sample which is limited using 
bright and faint limiting magnitudes in $G$-band. Relative distance errors 
$\sigma_{d_{\rm BJ}}/{d_{\rm BJ}}$ are also shown in the lowest panel.}\label{Fig3}
\end{center}
\end{figure}

%FIGURE 5
\begin{figure}
\begin{center}
\includegraphics[scale=0.9]{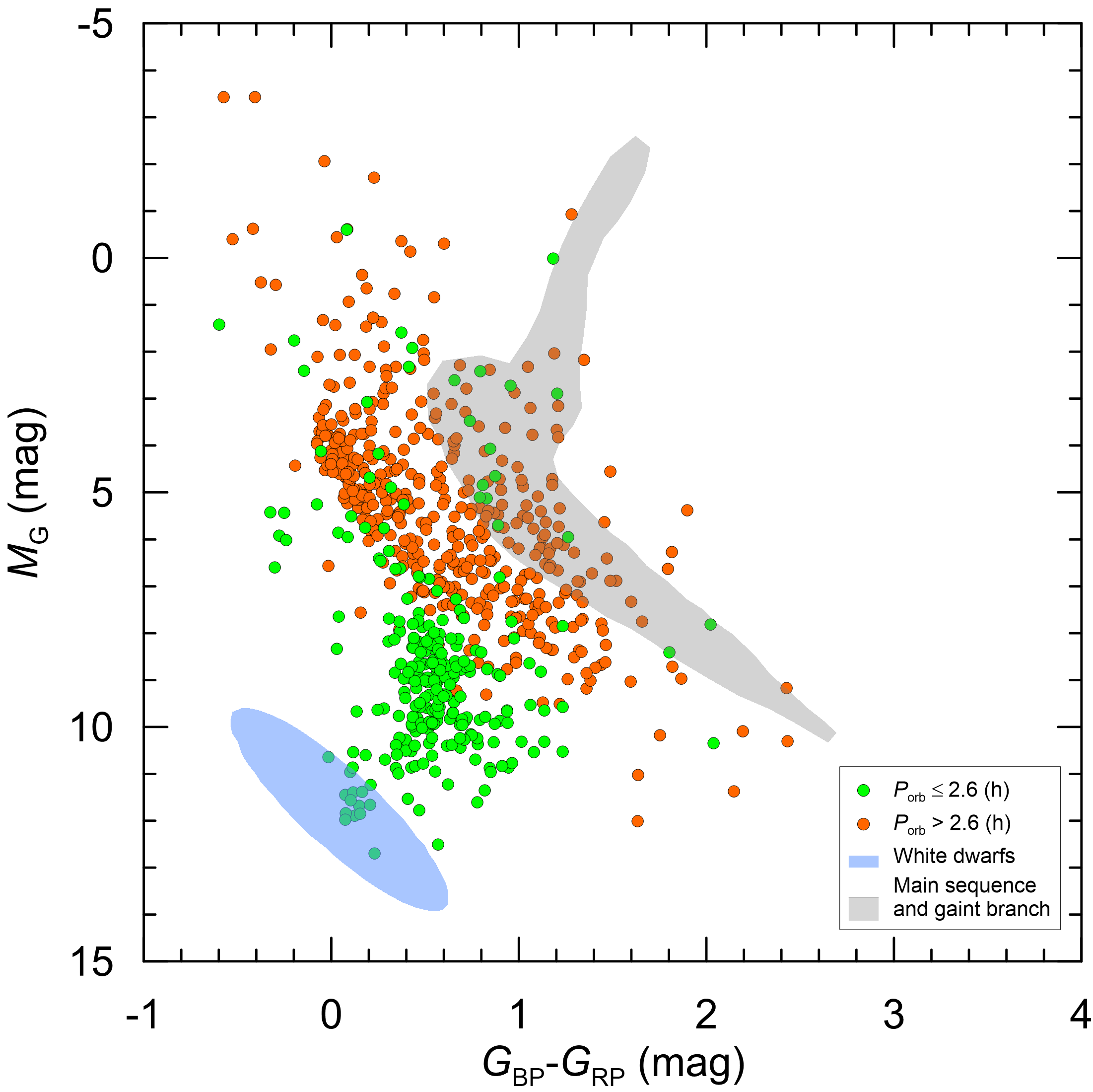}
\caption{{\it Gaia} DR3 HR diagram of CVs for which orbital periods $P_{\rm orb}$ 
are known in the final sample. The separating point of the orbital period gap is 
accepted as 2.6 h \citep{Ak10}. White dwarf, main-sequence and giant star regions 
are shaded in blue and grey, respectively \citep{Abretal20}. \label{Fig4}}
\end{center}
\end{figure}

%FIGURE 6
\begin{figure}
\begin{center}
\includegraphics[scale=0.08]{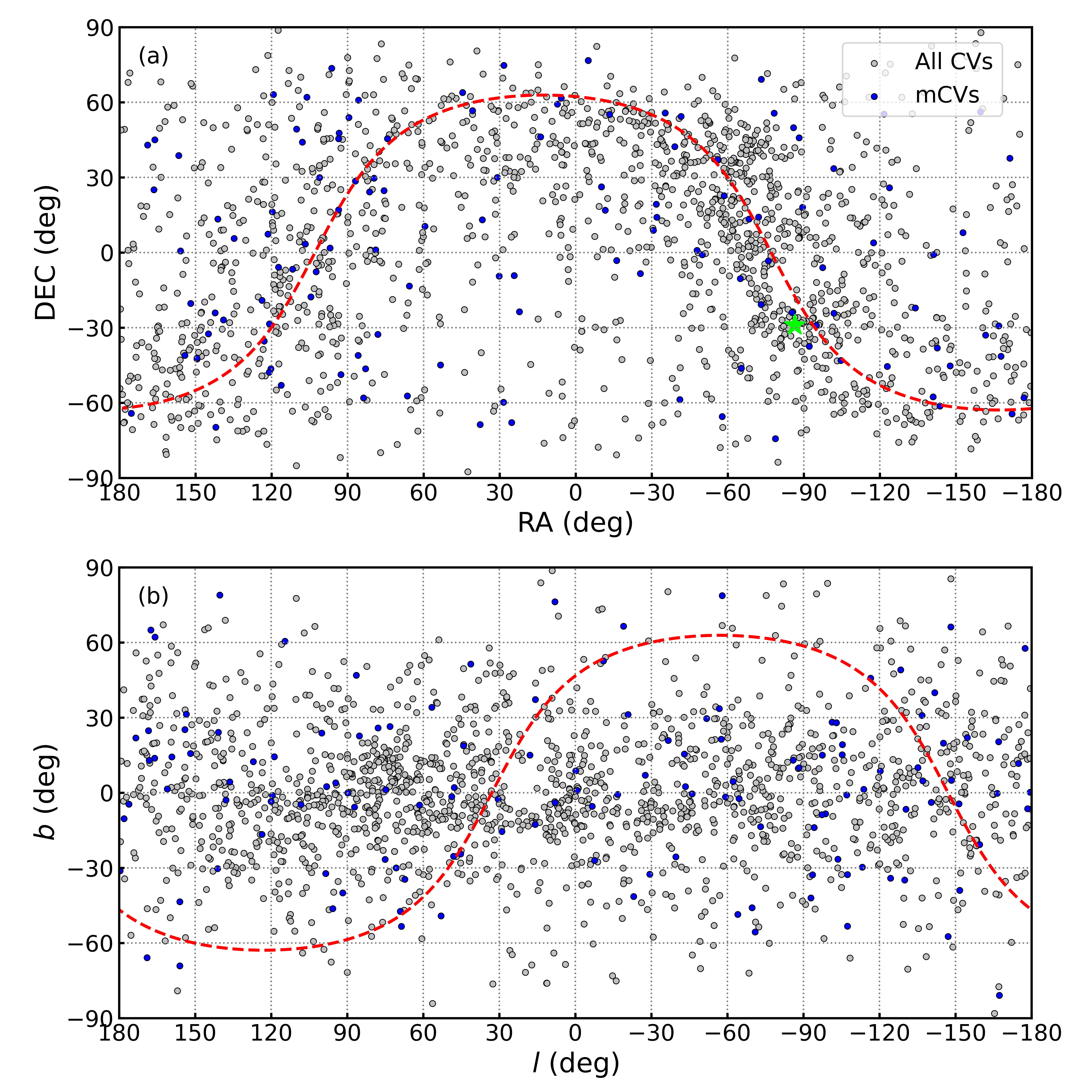}
\caption{The distribution of CVs according to an equatorial ($\alpha$, 
$\delta$) and Galactic ($l$, $b$) coordinates. Distributions of magnetic 
and non-magnetic systems are shown with different colours, blue and black, 
respectively. Red dashed lines represent the Galactic plane (upper panel) 
and the celestial equator (lower panel). The green star in the upper panel 
indicates the location of the Galactic centre on the plot.}
\end{center}
\end{figure}

Precision of distance estimations can be also seen in the HR diagram of CVs, for 
which the orbital periods are known. Figure 5 shows HR diagram of CVs below and 
above the orbital period gap using {\it Gaia} photometry. The separating point of 
the orbital period gap is accepted as 2.6 h \citep{Ak10}. CVs above the gap are 
generally brighter than those located below, as expected. This discrimination 
is very clear now due to precise distance measurements. It is clear that colour 
range for systems with $P_{\rm orb} \leq 2.6$ h is narrower compared to that of 
systems with $P_{\rm orb}>2.6$ h. Figure 5 shows that there are short-period systems brighter than $M_{\rm G}\approx 5$ mag. These are probably SU UMa type dwarf novae, which were in the superoutburst phase when they were observed. Blue and grey shaded parts in the HR diagram of CVs represent the regions where the white dwarf, main sequence and giant stars are located, respectively. The data for these shaded regions were taken from \citet{Abretal20}.  

\section{The Analysis and comparisons}
\subsection{Spatial distribution}

The distribution of CVs in the final sample according to equatorial and Galactic 
coordinates are plotted in Figure 6, upper and lower panels, respectively. The 
panels in Figure 6 indicate that the CVs are symmetrically distributed about the 
Galactic plane in general. The densest regions in both panels correspond 
to the Solar vicinity. In order to inspect the Galactic distribution of CVs in the 
Solar neighbourhood, we also calculated the Sun-centred rectangular Galactic 
coordinates of CVs ($X$ towards Galactic centre, $Y$ Galactic rotation, $Z$ north 
Galactic pole) in the sample and displayed their projected positions on the 
Galactic plane ($X-Y$ plane) and on a plane perpendicular to it ($X-Z$ plane) in 
Figure 7. Median heliocentric rectangular Galactic coordinates ($X$, $Y$, $Z$) are 
80, 93 and -18 pc, respectively, for all systems in the sample and -51, -4 and 24 
pc for magnetic systems, respectively. These values are summarized in Table 2. 

%FIGURE 7
\begin{figure*}
\begin{center}
\includegraphics[scale=0.80]{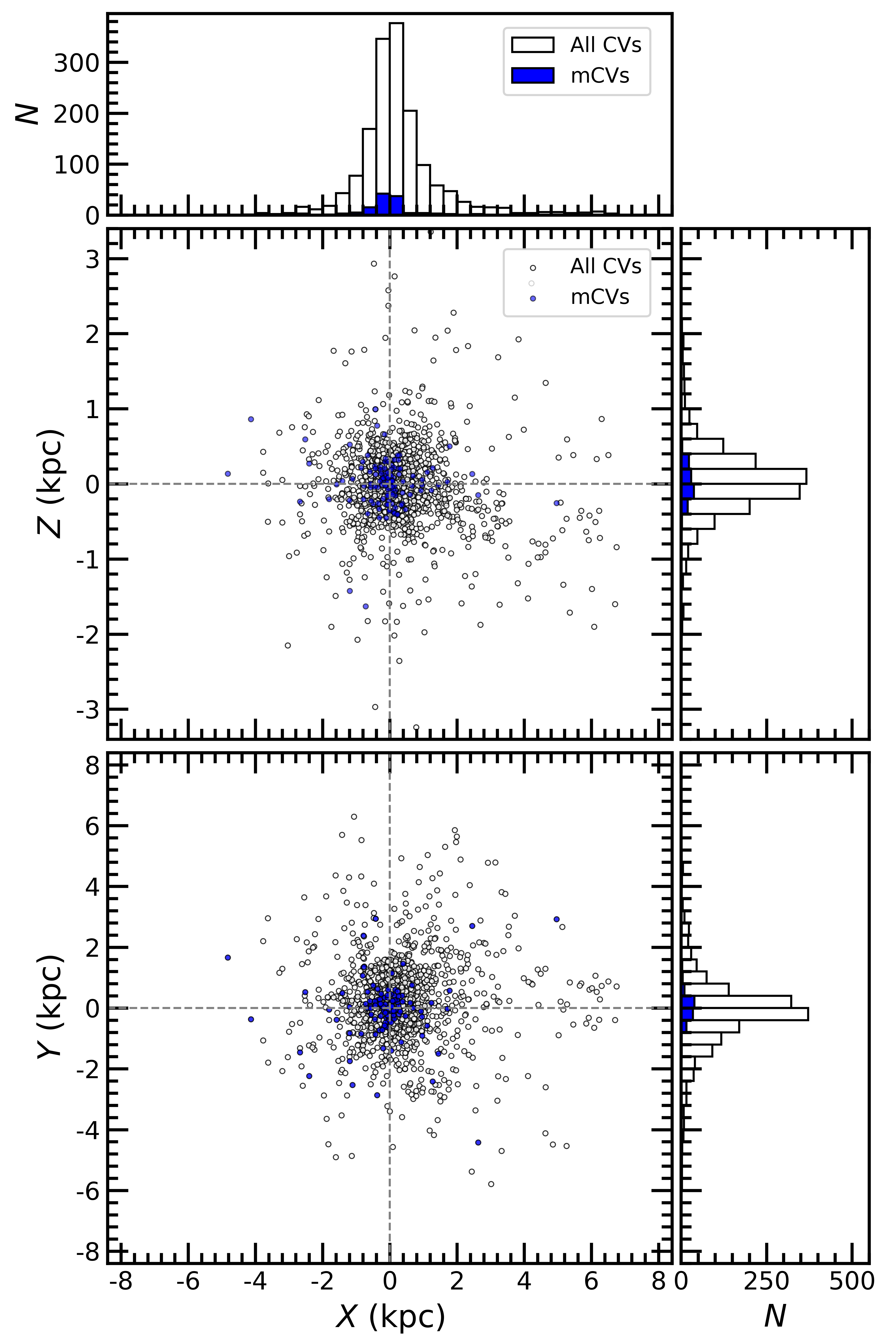}
\caption{The spatial distribution of CVs in the final sample with 
respect to the Sun. $X$, $Y$ and $Z$ are the Sun centred rectangular 
Galactic coordinates. Magnetic systems (mCVs) are shown with blue symbols. 
Number histograms created according to coordinates are shown to the right 
and top of the figure. }\label{Fig6}
\end{center}
\end{figure*}

%TABLE 2
\begin{table}
\normalsize
\begin{center}
\caption{The median distances ($d$) and heliocentric rectangular Galactic 
coordinates ($X$, $Y$, $Z$) of CVs in the sample. Values are separately listed 
for All CVs and magnetic (mCVs) systems. $N$ denotes the number of objects.}
\begin{tabular}{lccccc}
\hline 
\noalign{\vskip 0.1cm}
 Group      &   $N$  & $\tilde{d}$ & $\tilde{X}$  & $\tilde{Y}$  &  $\tilde{Z}$   \\ 
            &       & (pc) & (pc) & (pc) & (pc)   \\ 
 \hline
 All CVs        & 1587  & 989  & 80   &  93  & -18  \\ 
 %non-mCVs      & 1463  & 1021 & 102  & 105  & -22  \\ 
 mCVs           & ~124  &  559 & -51  &  -4  & ~~24 \\ 
 \hline
\end{tabular}  
\end{center}
\end{table}

Figures 6 and 7 show that there is no considerable bias according to the spatial 
distribution of CVs in our study. It should be noted that only magnetic CVs near 
the Sun can be detected as they are faint objects, in general. The median distances 
of CVs in the final sample are 989 and 559 pc for All CVs and magnetic systems, 
respectively. The median distances of All CVs and magnetic systems in the CV sample 
of \citet{Ozdn15} were found to be 423 and 385 pc, respectively. Comparisons of 
median distances in both studies and Figure 5 in this study with Figure 6 
of \citet{Ozdn15} reveals that much farther objects are included in our sample, 
thanks to {\it Gaia} mission. In addition, the number of CVs with distance estimates 
in this study is higher, and the distances are more accurate than 
those in \citet{Ozdn15}.  

\subsection{Galactic model parameters}

Using the number of objects per unit volume, it is possible to obtain 
information on their Galactic population. Galactic locations of objects 
must be known in order to estimate their Galactic model parameters. This
information allows us to estimate scale length and scale height of the 
objects in question. Based on deep sky surveys, it is known that the 
scale length of the thin disc stars is expected to be larger than 2.6 kpc 
\citep{Biletal06a,Juretal08}. Besides All CVs being members of the 
thin-disc population of the Galaxy according to their Galactic 
kinematics \citep{Ak15}, most of the systems in our sample are located in 
distances less than 2 kpc, as shown in Figure 7, and the median 
distance of All CVs in the final sample is 989 pc, which is much 
less than 2.6 kpc. Thus, our sample was not inspected according to 
population types and the scale length estimation was not performed. 

In order to find the Galactic model parameters of CVs, $z$-histograms that 
demonstrate the vertical distribution of objects in the Galaxy must be 
studied. Although the Galactic model parameters are derived using the 
exponential functions, \citet{Biletal06a} showed that the observed 
vertical distribution in the Galaxy is smoother in the Solar neighbourhood, 
and is well-approximated by a secans hiperbolicus square function 
(${\rm sech}^{2}$). Thus, the number of stars at a distance $z$ from the 
Galactic plane is described in our study by using both exponential and 
secans hiperbolicus square functions

\begin{equation}
n(z)=n_{0}\exp\Biggl(-\frac{\mid z\mid}{H}\Biggr) 
\end{equation}

and 

\begin{equation}
n(z)=n_{0}~{\rm sech}^{2}\Biggl(-\frac{\mid z\mid}{H_{z}}\Biggr),
\end{equation}
respectively. $n(z)$ based on ${\rm sech}^{2}$ can be also expressed as 

\begin{equation}
n(z)=n_{0}~\Biggl(\frac{4}{\exp{(-2z/H_{z})}+\exp{(2z/H_{z})+2}}\Biggr)
\end{equation}
\citep{Biletal06a}. Here, $z$ is the distance of objects from the Galactic plane 
and $n_{0}$ is the number of stars for $z=0$ pc. $H$ and $H_{z}$ are the 
exponential and ${\rm sech}^{2}$ scale heights, respectively. $z$ is described as
$z=z_{0}+d\sin(b)$, with $b$ being the Galactic latitude of the star, $d$ 
distance of the object and $z_{0}$ distance of the Sun from the Galactic 
plane \citep[24 pc;][]{Juretal08}. The relation between the exponential scale 
height $H$ and the ${\rm sech}^{2}$ scale height $H_{z}$ is 
$H=1.08504\times H_{z}$ \citep{Biletal06a}. To sample the posteriors, we used 
the Markov Chain Monte Carlo (MCMC) \textit{emcee} package of the affine-invariant 
ensemble sampler \citep{GW10}, kindly provided by \citet{Foretal13}. To obtain 
initial model parameters, we used a nonlinear least-squares algorithm (Python 
library LMFIT). Using these priors, we ran the MCMC to sample the posteriors 
for 128 initial conditions and tested these walkers in 15,000 steps of chains. 
Thus, we obtained the most plausible model parameters and their errors through 
minimising chi-square ($\chi_{\rm min}^{2}$).

The best fits to the $z$-histograms of All CVs and magnetic CVs in the 
sample are shown in Figure 8. We also demonstrate the 2-D posterior probability 
distributions of the model parameters sampled by MCMC in Figure 8. The scale 
height $H$ and the number of stars in the Solar neighbourhood $n_{0}$ obtained 
from the analyses are listed in Table 3. 

%FIGURE 08
\begin{figure*}
\begin{center}
\includegraphics[scale=0.58]{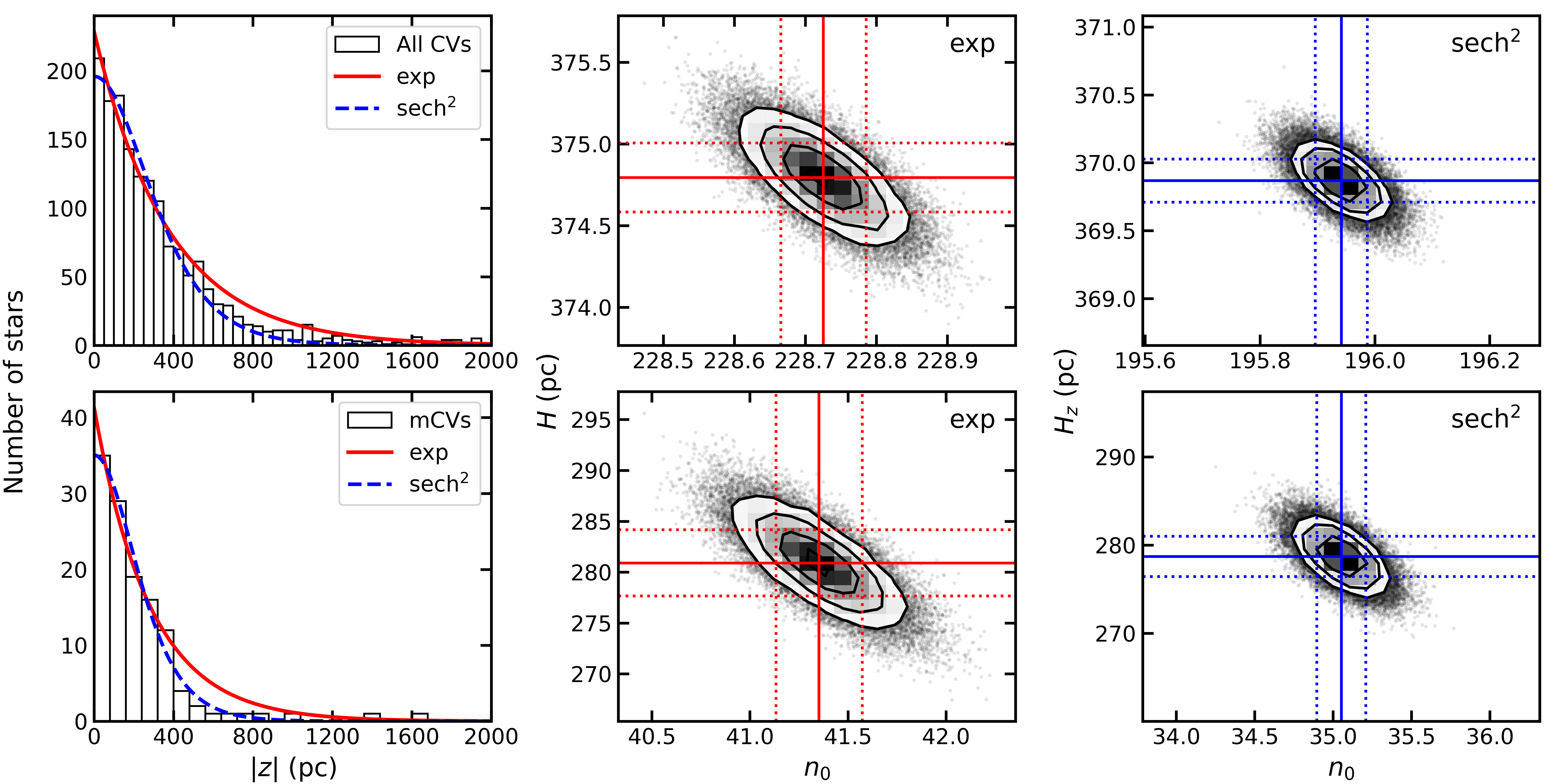}
\caption{The $z$-histograms for CVs. The upper panel shows the $z$-histogram 
for All CVs in the sample and the lower panel magnetic systems (mCVs). The blue 
dashed line represents the ${\rm sech}^{2}$ function and the red solid line 
the exponential function. The 2-D posterior probability distributions of the 
model parameters sampled by MCMC are demonstrated to the right of the 
$z$-histograms.}\label{Fig7} 
\end{center}
\end{figure*}

%TABLE3
\begin{table}
\setlength{\tabcolsep}{8pt}
\normalsize
\begin{center}
\caption{The Galactic model parameters for All CVs and mCVs in the sample. Model 
functions are given in the third column. Here, exp 
denotes for exponential function and ${\rm sech}^{2}$ for secans 
hiperbolicus square function. $N$ is the number of systems in the object 
group, $n_{0}$ is the number of stars in the Solar neighbourhood, $H$ the 
scale height for the model function.}
\begin{tabular}{lcccc}
\hline
 Group      & $N$  & Function & $n_{0}$  & $H$(pc)\\ 
 \hline
 All CVs    & 1587  &    exp           & 229$\pm$1 & 375$\pm$2\\ 
            &       & ${\rm sech}^{2}$ & 196$\pm$1 & 370$\pm$1\\ 
  \hline
  mCVs      &  124  &    exp           & 41$\pm$1  & 281$\pm$3\\
            &       & ${\rm sech}^{2}$ & 35$\pm$1  & 279$\pm$3\\ 
 \hline
\end{tabular}  
\end{center}
\end{table}

%FIGURE 09
\begin{figure*}
\begin{center}
\includegraphics[scale=0.57]{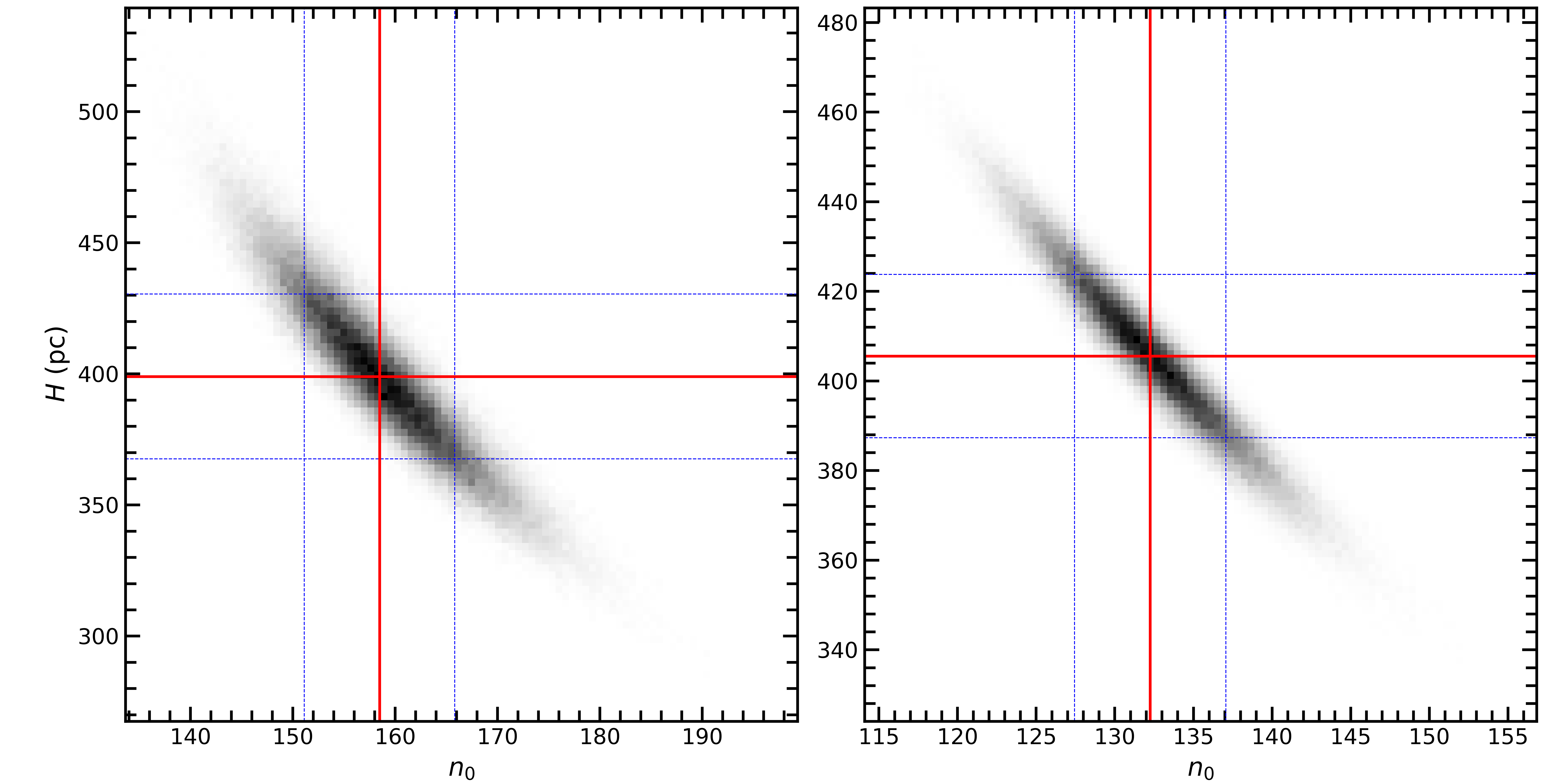}
\caption{The number of stars for $z=0$ pc ($n_{0}$) versus scale height ($H$), 
estimated with 15,000 trials for the Monte Carlo simulations. The simulations 
were performed for All CVs in the sample, keeping magnetic systems in the sample 
as they are and assuming 22$\%$ of the remaining sample to be magnetic by 
arbitrary selection. The left and right panels are plotted for exponential and 
${\rm sech}^{2}$ scale heights, respectively. Red straight lines show the most 
probable values. 1$\sigma$ values are indicated with blue straight lines.}\label{Fig8}
\end{center}
\end{figure*}

It is remarkable that exponential and ${\rm sech}^{2}$ functions give very 
similar scale heights for the two groups in Table 3. Nevertheless, the 
exponential functions well represent the $z$-histograms in Figure 8 
constructed for All CVs and magnetic CVs in the sample. From Table 3, we found 
that there is a considerable difference between the scale heights obtained 
for All CVs and magnetic systems, 375$\pm$2 and 281$\pm$3 pc, respectively. 
This can be expected as the median distances of these two groups of systems 
are 989 and 559 pc, respectively. Thus, we conclude that magnetic CVs are 
different than All CVs in the sample with respect to the scale height. 
In addition, as a consequence of reduced magnetic braking with the strong 
magnetic fields \citep{LWW94}, magnetic CVs evolve slower than non-magnetic 
systems \citep{Arau05}. Thus, we expect that their Galactic parameters 
should be different. \cite{Ak13} estimated the contribution of thick-disc CVs 
in the Solar neighbourhood to the Galactic model parameters from the Monte Carlo 
simulations and found that only about 6 per cent of CVs in the Solar 
neighbourhood are members of the thick-disc population of the Galaxy. Therefore, 
the effect of thick disc systems on the scale heights in Table 3 must be negligible. 

Another interesting finding is that, if the $z$ histogram of CVs is accepted 
to be exponential, it must be about 25 missing CVs within the sphere with 
a radius of about 100 pc with the Sun at the centre, which reminds that the number of 
period bouncers discovered in sky surveys is less than expected from 
the population models based on the standard theory. \citet{McAl19} found that 
30\% of donor stars in their sample are likely to be brown dwarfs 
in period bouncers, while only 5\% of CVs are located within 150 pc from the Sun 
in \citet{Pala20} included period bouncers. 

Although we selected magnetic systems (AM Her or DQ Her type systems) according 
to their classification in the literature, in fact, it can be unclassified magnetic 
systems in the sample. In a sense, the sample could be contaminated by them. 
As a result the scale heights estimated for All CVs in Table 3 could not be 
reliable. In order to find the effect of magnetic system contamination on the 
scale height of All CVs, we decided to perform Monte Carlo simulations. An 
inspection shows that about 30$\%$ of CVs in the catalogue of \citet{RK03} are 
classified as either AM Her or DQ Her system. Thus, we assumed that 30$\%$ of 
the systems in our sample is magnetic. As we know that 124 of CVs in the sample 
are classified as magnetic, which is about 8$\%$ of objects in this study. Keeping 
magnetic systems in the sample as they are, 22$\%$ of the remaining sample were 
assumed to be magnetic by arbitrary selection. We calculated $n_{0}$ and $H$ for 
each run of 15,000 trials for the Monte Carlo simulations. Figure 9 shows 
$n_{0}$ versus $H$ for the logarithmic and ${\rm sech}^{2}$ functions. 
After 15,000 trials for the Monte Carlo simulations performed on All CVs 
in the sample, the most probable values of the scale heights for the logarithmic 
and ${\rm sech}^{2}$ functions were found to be 398$\pm$31 and 406$\pm$18 pc, 
respectively. Error values are 1$\sigma$ errors. A comparison with the 
values in Table 3 shows that these values are in agreement, within errors, 
with those given for All CVs in the sample and the effect of the magnetic system' 
contamination on the scale height of all systems can be negligible for our sample. 
Note that the most probable $n_{0}$ values obtained from the Monte Carlo 
simulations are 158$\pm$7 and 132$\pm$5 for the logarithmic and ${\rm sech}^{2}$ 
functions, respectively. Differences of these values from those in Table 3 are 
expected as the system numbers in different $z$ distances were changed during 
the simulations.

A comparison of the scale heights found in this study with the ones 
in \citet{Ozdn15} reveals that their results are very different from those listed 
in Table 3. They found the exponential scale heights to be 213$^{+11}_{-10}$ and 
173$^{+18}_{-15}$ pc for All CVs and magnetic systems in their sample, respectively. 
It is clear that these values are very different than 375$\pm$2 and 
281$\pm$3 pc in Table 3. The scale height of about 375 pc found for All CVs 
in this study is also considerably larger than those suggested by 
\citet[][190$\pm$30 pc]{Pat84}, \citet*[][160-230 pc]{vP96} and 
\citet[][158$\pm$14 pc]{Ak08}. Note that \citet{Pala20} assumed a scale height 
of 280 pc for their analysis. 

The scale height differences between this study and the previous estimates should 
be resulted from the number of systems in the analysis and the accuracy of distance 
estimates. As our data sample is based on reliable distances based on precise  
trigonometric parallax measurements and it includes the highest number of systems 
in a similar analysis in the literature, we believe that the results in Table 3 
can be confidently used in population studies and further analysis of CVs.

We also derived the scale height of CVs in terms of the orbital period to 
find if this Galactic model parameter changes according to the orbital period. 
In order to compare our results with those in \citet{Ozdn15}, the CVs in our 
sample were divided into four-period intervals as defined in their study. Thus, 
there are 203 systems in the period interval $1.37 \leq P_{\rm orb}$(h) $< 2.25$, 
157 systems in $2.25 \leq P_{\rm orb}$(h) $< 3.7$, 105 systems in 
$3.7 \leq P_{\rm orb}$(h) $< 4.6$ and 171 in $4.6 \leq P_{\rm orb}$(h) $< 12$.
The best fits to the $z$-histograms of CVs grouped according to these 
period intervals are shown in Figure 10. The scale height $H$ and the 
number of stars in the Solar neighbourhood $n_{0}$ obtained from the 
minimum $\chi^{2}$ analysis are listed in Table 4 for the period intervals 
given above. Since the number of systems is too small for magnetic CVs in these  
period ranges, we list the Galactic model parameters only for All CVs in the sample. 

The $z$-histograms in Figure 10  are well represented by exponential 
functions, in general. A comparison of scale heights in Table 4 with those 
given in \citet{Ozdn15} shows the same trend, although the values in this 
study are more reliable. The scale height increases monotonously from 
248$\pm$2 to 430$\pm$4 pc while the orbital period decreases from 12 to 2.25 h. 
However, it drops to 300$\pm$2 pc for the shortest orbital period CVs with 
$P_{\rm orb} < 2.25$ h. We found a similar trend also for the ${\rm sech}^{2}$ 
function. The scale height for the systems in the interval 
$2.25 \leq P_{\rm orb}{\rm (h)}<3.70$ is the highest for all CVs. Note that 
only $\sim14\%$ of the systems in this interval are classified as mCV in 
our sample.  

%FIGURE 10
\begin{figure*}
\begin{center}
\includegraphics[scale=0.70]{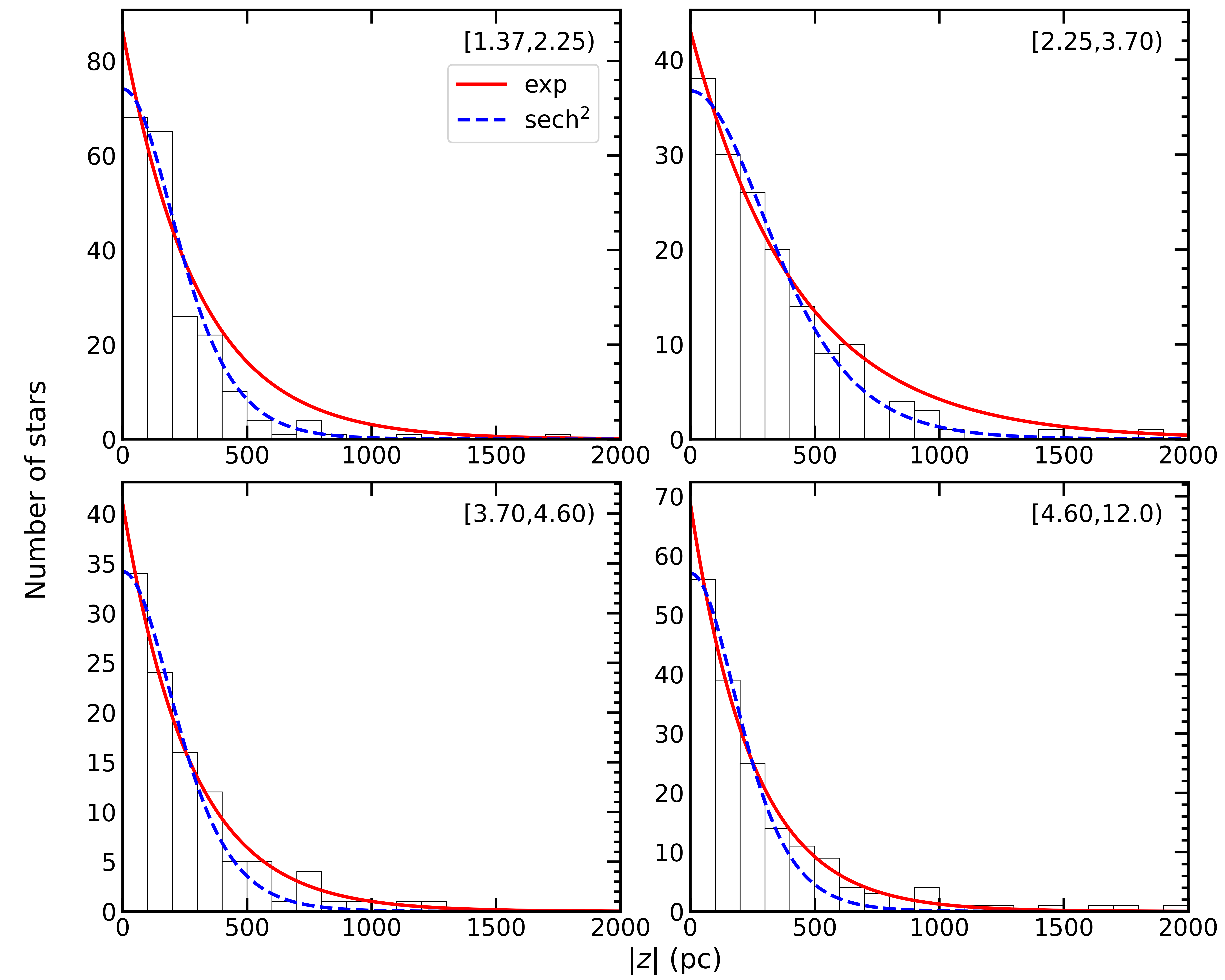}
\caption{The $z$-histograms for All CVs in the sample in terms of the orbital 
period. Orbital period ($P_{\rm orb}(\rm h)$) intervals are shown in brackets. 
The best fits to the $z$-histograms are also presented. The blue dashed line 
represents the ${\rm sech}^{2}$ function and the red solid line the exponential 
function.}\label{Fig9}
\end{center}
\end{figure*}

%TABLE 4
\begin{table*}
\normalsize
\begin{center}
\caption{The Galactic model parameters for All CVs in the sample in terms 
of the orbital period $P_{\rm orb}(\rm h)$. $n_{\rm 0}$ and $H$ are as 
defined in Table 3.}
\begin{tabular}{lcccc}
\hline
$P_{\rm orb}(\rm h)$  & $N$  &  Function  & $n_{0}$  & $H$ (pc)     \\ 
 \hline
[1.37 , 2.25)         & 203 & $\exp$           & 86$\pm$1 & 300$\pm$2 \\ 
                      &     & ${\rm sech}^{2}$ & 74$\pm$1 & 286$\pm$1 \\ 
\hline
[2.25 , 3.70)         & 157 & $\exp$           & 43$\pm$1 & 430$\pm$4  \\
                      &     & ${\rm sech}^{2}$ & 37$\pm$1 & 424$\pm$3  \\ 
\hline
  [3.70 , 4.60)       & 105 &    $\exp$        & 41$\pm$1 & 269$\pm$4  \\
                      &     & ${\rm sech}^{2}$ & 34$\pm$1 & 278$\pm$3  \\ 
\hline
  [4.60 , 12.00)      & 171 &   $\exp$         & 69$\pm$1 & 248$\pm$2  \\
                      &     & ${\rm sech}^{2}$ & 57$\pm$1 & 258$\pm$2  \\ 
\hline 
\end{tabular}  
\end{center}
\end{table*}

\subsection{Space density}

Space density is an important parameter for population synthesis studies 
based on theoretical evolutionary models of a selected object type. The space 
density of a group of stars is derived by dividing the number of stars 
in consecutive distances from the Sun to the corresponding partial spherical 
volumes: $D=N/\Delta V_{i, i+1}$ \citep{Biletal06a,Biletal06b,Biletal06c}. Here, $D$ is 
the space density, $N$ denotes the number of stars in the partial spherical 
volume $\Delta V_{i, i+1}$ which is defined by consecutive distances $d_i$ and 
$d_{i+1}$ from the Sun. The logarithmic space density is preferred to compare 
the results in the literature, which is defined as $D^{*}=\log D + 10 $. The 
logarithmic density functions of All CVs and magnetic systems in the Solar 
neighbourhood are shown in Figure 11, where $r^{*}$ denotes the centroid 
distance of the partial spherical volume which is defined 
as $r^{*}=[(d^{3}_{i}+d^{3}_{i+1})/2]^{1/3}$. The local space density $D_0$ is 
the space density estimated for $r^{*}=$ 0 pc from the exponential fits shown 
in Figure 11. The logarithmic and local space densities of CV groups in 
the Solar neighbourhood are listed in Table 5.

%TABLE 5
\begin{table}
\normalsize
\begin{center}
\caption{The logarithmic and local space densities of CVs. Symbols for subgroups are 
as in Table 3. $N$ denotes the number of stars in the subgroup, $D_0$ is the local 
space density and $D^{*}$ logarithmic space density.}
\begin{tabular}{lccc}
\hline
 Group      &  $N$  &   $D^{*}$     &           $D_0$              \\ 
            &       &               & ($\times$10$^{-6}$ pc$^{-3}$) \\ 
  \hline
 All CVs    & 1587  & 4.83$\pm$0.07 & $6.8^{+1.3}_{-1.1}$        \\ 
   %\hline
   \noalign{\vskip 0.2cm} 
  mCVs      &  124   & 4.33$\pm$0.09 & $2.1^{+0.5}_{-0.4}$        \\
  \hline
\end{tabular}  
\end{center}
\end{table}

%FIGURE 11
\begin{figure}
\begin{center}
\includegraphics[scale=0.75]{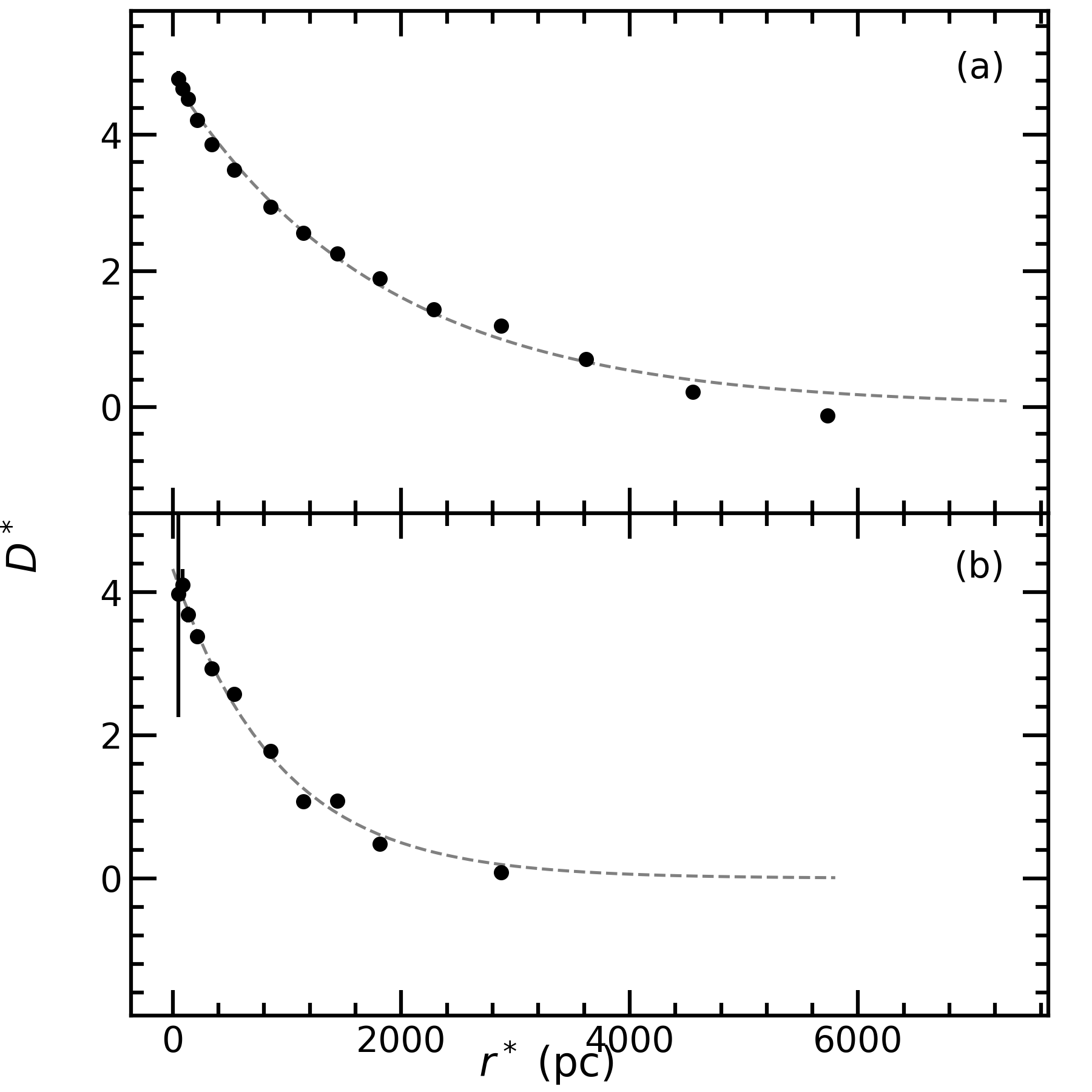}
\caption{The logarithmic density functions of All CVs (panel a) and mCVs (panel b) 
in the sample. Dashed lines represent exponential fits applied to the data.}\label{Fig10}
\end{center}
\end{figure}

Table 5 shows that the local space density of All CVs in the sample is 
$6.8^{+1.3}_{-1.1}\times 10^{-6}$ pc$^{-3}$. The space density of magnetic 
CVs are about three times smaller than that found for all systems, 
$2.1^{+0.5}_{-0.4}\times 10^{-6}$ pc$^{-3}$. The local space density 
estimation in this study takes into account the CVs located even further 
than 6 kpc, which corresponds to a large Galactic volume. Note that the 
median distance of the objects in the sample is 989 pc and the  
local space density estimation is based on the objects within the 
completeness limits. Therefore, we believe that the space density values 
obtained from the CV sample in this study are reliable. 

CV population synthesis models based on the standard formation and evolution 
scenario predicts space densities $10^{-5}$-$10^{-4}$ pc$^{-3}$ 
\citep{RB86,deKool92,Kolb93,Pol96,Wil05,Wil07,GN15,Bel18} while the space densities 
found in observational studies are on the order of $10^{-7}$-$10^{-4}$ pc$^{-3}$  
\citep{War74,Pat84,Pat98,TB98,Rin93,Schw02,Arau05,Pret07,PKK07,Ak08,Revn08,PK12,PKS13,Schw18}
In a recent study, \citet{Pala20} measured the space density of  
$4.8^{+0.6}_{-0.9}\times10^{-6}$ and $1.2^{+0.4}_{-0.5}\times10^{-6}$ pc$^{-3}$ 
for All CVs and mCVs, respectively, from {\it Gaia} DR2 \citep{Gaia18}. 
They assumed a scale height of 280 pc for their analysis and their 
data sample included 42 objects within 150 pc from the Sun. Note that we 
found a scale height of $H=375$ pc for All CVs (see Table 3).

The local space density estimated for All CVs in our sample 
($6.8^{+1.3}_{-1.1}\times 10^{-6}$ pc$^{-3}$) is 2-20 times smaller than 
those predicted by population synthesis studies based on the standard evolution 
scenario. However, it is very similar to the observational space density 
found by \citet{Pala20} ($4.8^{+0.6}_{-0.9}\times10^{-6}$ pc$^{-3}$) from 
{\it Gaia} DR2 \citep{Gaia18}, within the errors. The local space densities 
obtained in this study are only about 1.5 times more than those they found. 
Note that the fraction of mCVs in the whole sample used in this study is about 8$\%$.

\subsection{Luminosity function}

The luminosity function is defined as the space density of objects 
in a certain absolute magnitude interval \citep{Karaali03, Karaali04, Karaali09, Ak07}. 
We estimated the logarithmic luminosity functions $\phi$ for All CVs and 
magnetic systems and presented them in Table 6, where $\Delta V_{i,i+1}$ is the 
partial spherical volume which includes the objects located between the distances 
$d_i$ and $d_{i+1}$. These distance limits correspond to the bright and faint 
limiting apparent magnitudes in $G$-band, 9 $\leq G \leq$ 18.5 mag, for the absolute 
magnitude $M_{\rm G}$ interval preferred here. The logarithmic luminosity functions 
of all systems and magnetic systems are plotted in the lower panel of Figure 12. As can 
be expected, the luminosity function of All CVs is considerably different than the 
one estimated for magnetic CVs.

%TABLE 6
\begin{table*}
\small
\center
\caption{The logarithmic luminosity functions $\phi$ of CVs in the sample with 
9 $\leq G \leq$ 18.5 mag. $N$ is the number of stars in the $M_{\rm G}$ absolute 
magnitude interval given in the first column. $\Delta V_{i,i+1}$ is the partial 
spherical volume that includes the objects between the distances $d_i$ and $d_{i+1}$ 
corresponding to the bright and faint limits in $G$-band for the absolute magnitude  
$M_{\rm G}$ interval. $\rho$ denotes the density and $r^{*}$ the centroid distance 
of the partial spherical volume.}

\begin{tabular}{crrrrrrrrrr}
\hline
& & &  &  & \multicolumn{3}{c}{All CVs} & \multicolumn{3}{c}{mCVs}\\
\cline{6-11} 
$M_{{\rm G}}$  & \multicolumn{1}{c}{$d_{1}$} & \multicolumn{1}{c}{$d_{2}$}  & \multicolumn{1}{c}{$\Delta V_{1,2}$} & \multicolumn{1}{c}{$r^{*}$ } & \multicolumn{1}{c}{$N$} 
& \multicolumn{1}{c}{$\rho$}  & \multicolumn{1}{c}{$\phi$} & \multicolumn{1}{c}{$N$} & \multicolumn{1}{c}{$\rho$} & \multicolumn{1}{c}{$\phi$}   \\
\multicolumn{1}{c}{(mag)} & \multicolumn{1}{c}{(pc)} & \multicolumn{1}{c}{(pc)} & \multicolumn{1}{c}{($\rm pc^3$)} &  \multicolumn{1}{c}{(kpc)}   &  
& \multicolumn{1}{c}{($\rm pc^{-3}$)}  &         &   & \multicolumn{1}{c}{($\rm pc^{-3}$)}   &   \\
\hline
\noalign{\vskip 0.1cm}    
$[$-3,-2) & 2512 & 125893 & 8.4$\times10^{15}$ & 99.92 &  11 & 1.3$\times10^{-15}$ & -14.9 & 1 & 1.2$\times10^{-16}$  & -15.9   \\
$[$-2,-1) & 1585 &  79433 & 2.1$\times10^{15}$ & 63.05 &  8 & 3.8$\times10^{-15}$ & -14.4 & 1 & 4.8$\times10^{-16}$  & -15.3   \\ 
$[$-1, 0) & 1000 &  50119 & 5.3$\times10^{14}$ & 39.78 & 25 & 4.7$\times10^{-14}$ & -13.3 & 1 & 1.9$\times10^{-15}$  & -14.7   \\
$[$ 0, 1) &  631 &  31623 & 1.3$\times10^{14}$ & 25.10 & 22 & 1.7$\times10^{-13}$ & -12.8 &  5 & 3.8$\times10^{-14}$ & -13.4  \\ 
$[$ 1, 2) &  398 &  19953 & 3.3$\times10^{13}$ & 15.84 & 28 & 8.4$\times10^{-13}$ & -12.1 &  3 & 9.0$\times10^{-14}$ & -13.0  \\ 
$[$ 2, 3) &  251 &  12589 & 8.4$\times10^{12}$ &  9.99 & 87 & 1.0$\times10^{-11}$ & -11.0 &  5 & 6.0$\times10^{-13}$ & -12.2  \\
$[$ 3, 4) &  158 &   7943 & 2.1$\times10^{12}$ &  6.30 & 164 & 7.8$\times10^{-11}$ &  -10.1 &  10 & 4.8$\times10^{-12}$ & -11.3  \\ 
$[$ 4, 5) &  100 &   5012 & 5.3$\times10^{11}$ &  3.98 & 229 &  4.3$\times10^{-10}$ &  -9.4 & 14 & 2.7$\times10^{-11}$ & -10.6   \\ 
$[$ 5, 6) &   63 &   3162 & 1.3$\times10^{11}$ &  2.51 & 275 &  2.1$\times10^{-09}$ &  -8.7 &  12 & 9.1$\times10^{-11}$ & -10.0   \\ 
$[$ 6, 7) &   40 &   1995 & 3.3$\times10^{10}$ &  1.58 & 210 &  6.3$\times10^{-09}$ &  -8.2 &  8 & 2.4$\times10^{-10}$ &  -9.6   \\ 
$[$ 7, 8) &   25 &   1259 & 8.4$\times10^{09}$ &  1.00 & 159 &  1.9$\times10^{-08}$ &  -7.7 &  13 &  1.6$\times10^{-09}$ &  -8.8   \\ 
$[$ 8, 9) &   16 &    794 & 2.1$\times10^{09}$ &  0.63 & 159 &  7.6$\times10^{-08}$ &  -7.1 & 24 &  1.1$\times10^{-08}$ &  -7.9   \\ 
$[$ 9,10) &   10 &    501 & 5.3$\times10^{08}$ &  0.40 &  99 &  1.9$\times10^{-07}$ &  -6.7 &  10 &  1.9$\times10^{-08}$ &  -7.7   \\ 
$[$10,11) &    6 &    316 & 1.3$\times10^{08}$ &  0.25 &  65 &  4.9$\times10^{-07}$ &  -6.3 &  10 &  7.6$\times10^{-08}$ &  -7.1   \\       
$[$11,12) &    4 &    200 & 3.4$\times10^{07}$ &  0.16 &  28 &  8.4$\times10^{-07}$ &  -6.1 &  6 &  1.8$\times10^{-07}$ &  -6.7   \\     
$[$12,13) &    3 &    126 & 8.4$\times10^{06}$ &  0.10 &   7 &  8.4$\times10^{-07}$ &  -6.1 & 1 & 1.2$\times10^{-07}$  & -6.9   \\            
\hline
\end{tabular}
\end{table*}

%FIGURE 12
\begin{figure}
\begin{center}
\includegraphics[scale=0.80]{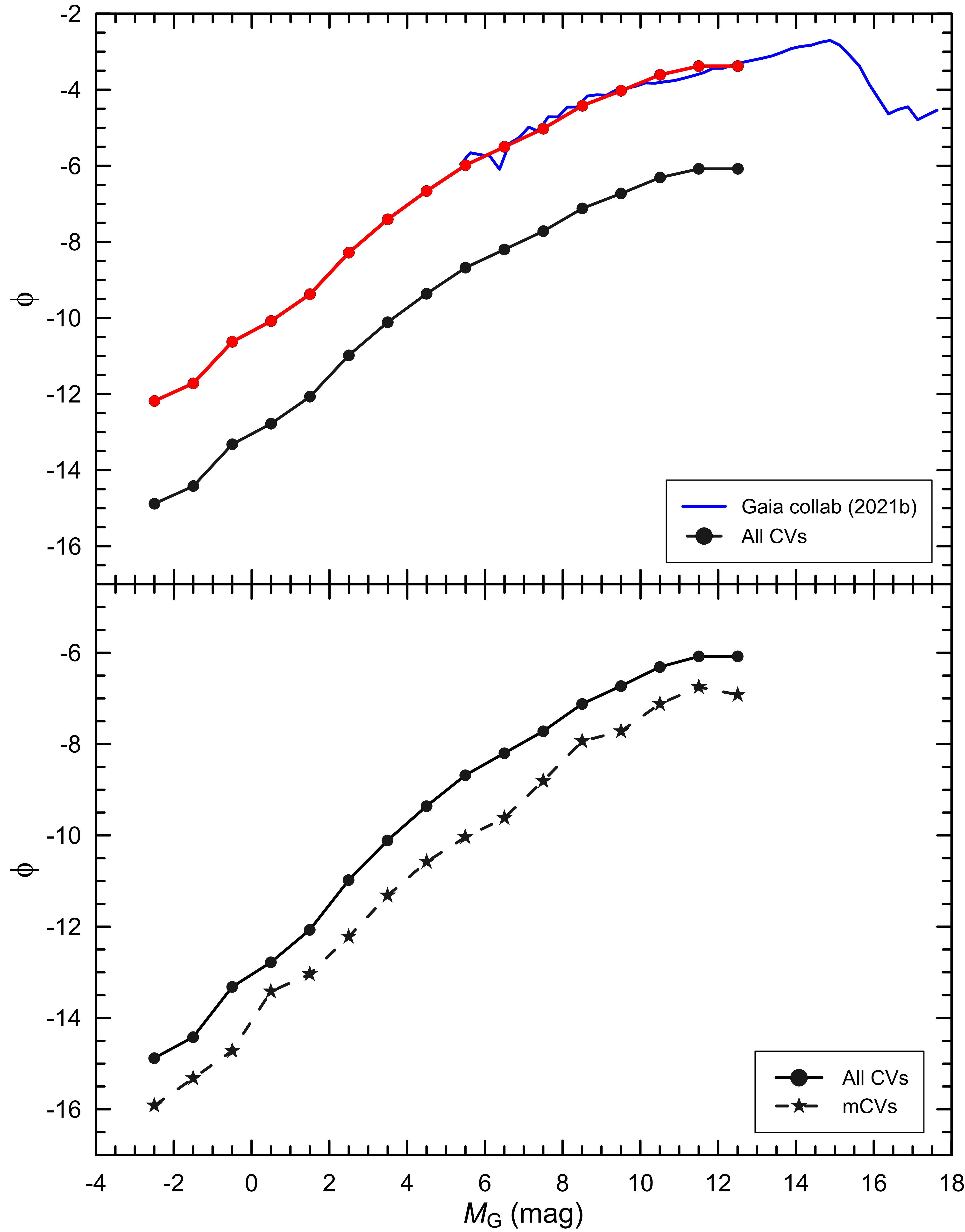}
\caption{Logarithmic luminosity functions of CVs. Lower panel shows luminosity 
functions for All CVs and mCVs. Upper panel 
demonstrates the logarithmic luminosity function for All CVs in the sample. 
Red line represent the 500 times the luminosity function of All CVs and 
the blue line logarithmic luminosity function of white dwarfs taken 
from \citet{Gaia21b}.}\label{Fig11}
\end{center}
\end{figure}

The luminosity function of All CVs in the data sample is plotted in the upper 
panel of Figure 12, where we also presented the white dwarf luminosity function 
that demonstrate the collective evolution of white dwarfs \citep{Gaia21b}. 
As can be seen from Table 6 and Figure 12, besides magnetic systems have 
a luminosity function smaller than that estimated for All CVs, they span
a narrower absolute magnitude interval compared to the absolute magnitude 
interval of all systems. The comparison of the luminosity functions of white 
dwarfs and All CVs in our sample reveals that the tendencies of both luminosity 
functions is almost the same and that the 500 times the luminosity function 
of All CVs corresponds to the continuation of the luminosity function of white 
dwarfs towards the brighter absolute magnitudes. A similar comparison was 
demonstrated by \citet{Ozdn15} for CVs in their data sample and white dwarfs 
in the Anglo Australian Telescope survey \citep{Boyle89} and the Palomar Green 
survey \citep*{Flemetal86}. 

The single white dwarf masses are significantly smaller than those of the white 
dwarfs in CVs \citep[see][and references therein]{ZS20} and it is claimed that 
the white dwarf mass does not depend on the orbital period \citep{McAl19,Pala22}. 
In addition, the mass of CV white dwarfs does not show a monotonous increase 
during the evolution of the system and the contribution of the primary component to 
the total radiation of a CV is dominant -or much higher- in UV rather than that in 
optical \citep{Gan00}. Simulations performed by \citet{Hill20} predicts that 
the white dwarf masses in CVs decrease monotonically, by only a few per cent 
throughout the evolution of cataclysmic variable. Therefore, it is unlikely that 
the trend of CV luminosity function in the upper panel of Figure 12 reflects the 
evolution of the white dwarf companion of these systems. In any case, this 
comparison shows that we find one CV for about 500 white dwarfs in the 
Solar neighbourhood.

\newpage

\section{Conclusions}

The spatial distribution, Galactic model parameters and luminosity function of 
cataclysmic variables were precisely derived using distances from \citet{Bai21}, 
who re-estimated trigonometric parallaxes of ESA's {\it Gaia} DR3 \citep{Gaia21a}. 
We compared distances obtained from \citet{Bai21} and {\it Gaia} DR3 data \citep{Gaia22}  
found that the scatter is too much for the systems $G\geq 18.5$ mag. Thus, the data sample 
in this study includes CVs with $9\leq G \leq 18.5$ mag. Number of CVs in the sample 
decreased from 10,852 to 1,587, with 124 of them are magnetic systems, by preventing the 
misidentification of CVs due to adjacent objects, checking for duplication and by also 
taking into account the quality flags of parallax measurements and completeness limits 
of the data.  

Projected positions of CVs on the Galactic plane ($X-Y$ plane) and on a plane 
perpendicular to it ($X-Z$ plane) demonstrate that systems in the sample are 
symmetrically distributed about the Galactic plane, in general. So, we conclude 
that there is no considerable bias according to the spatial distribution of CVs 
in our study. The median distances of objects in the sample are 989 and 559 for All 
CVs and magnetic systems, respectively.

The exponential scale heights were found to be 375$\pm$2 and 281$\pm$3 for 
All CVs and mCVs in the sample, respectively. Thus, we conclude that a scale 
height of 375 pc can be used in CV studies, in general. This value 
is considerably larger than those previously suggested in observational 
studies \citep{Pat84,vP96,Ak08}. It is also significantly larger than that  
estimated by \citet{Ozdn15}. Monte Carlo simulations showed that the magnetic 
system' effect on the scale height of all systems can be negligible 
for the sample.

It seems that it must be 25 missing CVs within the sphere with a radius 
of about 100 pc with the Sun at the centre. This reminds that the number of 
period bouncers discovered in sky surveys is less than expected from the population 
models based on the standard theory. Note that \citet{McAl19} found that 
30$\%$ of donor stars in their sample are likely to be brown dwarfs in period 
bouncers. 

The exponential scale heights of All CVs derived in terms of the orbital 
period shows that the scale height increases from 248$\pm$2 to 430$\pm$4 pc with the 
orbital period decreases from 12 to 2.25 h, and it almost suddenly drops to 
300$\pm$2 pc for the shortest orbital period CVs with $P_{\rm orb} < 2.25$ h. 
A similar trend was also found in \citet{Ozdn15}. Note that \citet{PKK07} modelled 
the Galactic population of CVs adopting 120, 260 and 450 pc for long,  
normal short orbital period systems and period bouncers, respectively. 

The local space density of All CVs and magnetic systems in the sample was 
estimated to be $6.8^{+1.3}_{-1.1}\times$10$^{-6}$ and $2.1^{+0.5}_{-0.4}\times10^{-6}$ 
pc$^{-3}$, respectively. The space densities estimated here for All CVs and magnetic 
systems are in agreement with those calculated by \citet{Pala20}, who used 42 CVs 
within 150 pc from the Sun with data obtained from {\it Gaia} DR2 \citep{Gaia18}, 
within errors. They measured the space densities of $4.8^{+0.6}_{-0.9}\times10^{-6}$ 
and $1.2^{+0.4}_{-0.5}\times10^{-6}$ pc$^{-3}$ for all CVs and magnetic CVs, respectively. 
We claim that the space density values in our study are the most reliable estimates 
ever found. The measurements in this study strengthen the discrepancy 
between CV space densities obtained from observations and those predicted by population 
synthesis models based on the standard formation and evolution theory of these systems. 
If the  population synthesis models are correct, this disagreement between the theory 
and observations means that the current CV surveys are incomplete as they missed almost all 
very low $\dot{M}$ systems and CVs in the period gap, through which the life 
time of a CV is predicted to be longer \citep{Hill20}, and period bouncers.

The logarithmic luminosity functions derived for CVs in the sample are in agreement with 
those shown in \citet{Ozdn15}. The trend of the logarithmic luminosity functions 
of CVs and white dwarfs are very similar. Although 500 times the luminosity function of CVs 
looks like the extension of the white dwarf luminosity function towards the brighter 
absolute magnitudes, it is not likely that this similarity indicates the evolution of 
white dwarf companion of CVs as the mass of CV white dwarfs does not show a monotonous 
increase during the evolution of the system and contribution of the primary component 
to the total radiation of a CV is dominant in UV \citep{Gan00,Pala22}. 

To conclude, the results in this study can be used in population studies and analysis of 
cataclysmic variables. We believe that the further $Gaia$ observations of cataclysmic 
variables and surveys focused on low $\dot{M}$ CVs and fainter systems will lead 
to not only larger datasets but also more precise distance measurements for these 
systems. Such observational results will allow us to obtain more detailed and confident 
observational Galactic model parameters to test population synthesis models. 

\section{Acknowledgments}
We would like to thank Michael Shara, the referee, for his useful and constructive 
comments concerning the manuscript. This work has been supported in part by the Scientific 
and Technological Research Council of Turkey (T\"UB\.ITAK) 119F072. This work has been 
supported in part by Istanbul University: Project number NAP-33768. This study is a part 
of the PhD thesis of Remziye Canbay. This work has made use of data from the European Space 
Agency (ESA) mission \emph{Gaia} (https://www.cosmos.esa.int/gaia), processed by 
the \emph{Gaia} Data Processing and Analysis Consortium 
(DPAC, https://www.cosmos.esa.int/web/gaia/dpac/ consortium). Funding for DPAC has been 
provided by national institutions, in particular, the institutions participating in 
the \emph{Gaia} Multilateral Agreement. This research has made use of NASA’s (National 
Aeronautics and Space Administration) Astrophysics Data System Bibliographic Services and 
the SIMBAD Astronomical Database, operated at CDS, Strasbourg, France and NASA/IPAC 
Infrared Science Archive and Extragalactic Database (NED), which is operated by the Jet 
Propulsion  Laboratory, California Institute of Technology, under contract with the 
National Aeronautics and Space Administration.

\bibliographystyle{pasa-mnras}
%\bibliography{1r_lamboo_notes}

\end{document}